\PassOptionsToPackage{pdfpagelabels=false}{hyperref}
\documentclass[fleqn,usenatbib]{mnras}

\usepackage[T1]{fontenc}
\usepackage{ae,aecompl}

\usepackage[pdftex]{graphicx}	
\usepackage{amsmath}	
\usepackage{amssymb}	
\usepackage{paralist}
\usepackage{diagbox}
\usepackage{float}
\usepackage{multicol}
\usepackage{tasks}
\usepackage{xr}
\usepackage{soul}
\usepackage[caption=false]{subfig}

\newcommand{\msol}{$M_{\odot} \, $}

\title[Multiwavelength emission of ULXs]{Modelling multiwavelength emission of Ultra-luminous X-ray Sources accreting above the Eddington limit}

\author[E. Ambrosi et al.]{
E. Ambrosi,$^{1}$\thanks{E-mail:elena.ambrosi@inaf.it }
L. Zampieri,$^{2}$
F. Pintore $^{1,3}$
A. Wolter $^{4}$
\\
$^{1}$\textit{INAF – IASF Palermo, Via U. La Malfa 153, 90146 Palermo, Italy;}\\
$^{2}$\textit{INAF-Osservatorio Astronomico di Padova, vicolo dell'Osservatorio,5, 35122 Padova, Italy;} \\
$^{3}$\textit{INAF - IASF Milano, via E. Bassini 15, I-20133 Milano, Italy;} \\
$^{4}$\textit{INAF, Osservatorio Astronomico di Brera, via Brera 28, 20121 Milano, Italy}
}

\date{Accepted XXX. Received YYY; in original form ZZZ}

\pubyear{2021}
\begin{document}
\label{firstpage}
\pagerange{\pageref{firstpage}--\pageref{lastpage}}
\maketitle

\begin{abstract}
We model the multiwavelength properties of binaries accreting at super-critical rates with the aim to better understand the observational properties of Ultra-luminous X-ray Sources (ULXs).
We calculate an extended grid of binary systems which evolve through Roche Lobe Overflow and undergo case A mass transfer from massive donors (up to 25 \msol) onto massive Black Holes (BH) (up to 100 \msol). Angular momentum loss with the ejection of mass through an outflow is incorporated. We apply our super-Eddington accretion model to these systems, computing their evolutionary tracks on the color-magnitude diagram (CMD) for the Johnson and HST photometric systems. We found that the tracks occupy specific positions on the CMD depending on the evolutionary stage of the donor and of the binary. Moreover, their shapes are similar, regardless the BH mass. More massive BHs lead to more luminous tracks. We additionally compute their optical-through-X-ray Spectral Energy Distribution (SED) considering the effects of a Comptonizing corona which surrounds the innermost regions of the disc. We apply our model to  four ULXs: NGC4559 X-7, NGC 5204 X-1, Holmberg II X-1 and NGC 5907 ULX-2. We found that accretion onto BHs with mass in the range $35-55$ \msol  is consistent with to the observational properties of these sources. We finally explore and discuss the possibility to extend our model also to ULXs powered by accreting Pulsars (PULXs).
\end{abstract}



%
\section{Introduction}
Ultra-luminous X-ray sources are point-like, off-nuclear X-ray sources whose bolometric luminosity exceeds the Eddington limit for a 10 \msol BH ($L_{Edd} \sim 2 \times 10^{39}$ erg s$^{-1}$; \citealt{1989ARA&A..27...87F}, see \citealt{2017ARA&A..55..303K} for a recent review). \\
They have been discovered in nearby galaxies and are associated with different environments: star-forming regions or young stellar environments in spiral galaxies, or dwarf irregular galaxies (e.g. \citealt{2006ApJ...641..241R}, \citealt{2006IAUS..230..293P}, \citealt{2007ApJ...661..165L}, \citealt{2012MNRAS.419.2095M}). In few cases they are associated with older stellar populations in elliptical galaxies (\citealt{2008ApJ...675.1067F}, \citealt{2008MNRAS.387...73R}). A large number of ULXs has been found also in the young and star forming rings of ring galaxies (\citealt{2018ApJ...863...43W}). Many of them are located in or close to young O-B associations (see e.g. the population studies performed by \citealt{2005MNRAS.356...12S,2008A&A...486..151G,2002MNRAS.335L..67G}), often embedded in low metallicity environments and/or high star-forming regions (\citealt{2011AN....332..414M,2015MNRAS.448..781W})\\ 
Their observational properties, as the X-ray luminosity, the variability and the spectral changes indicate that the majority ULXs are likely compact objects accreting via a disc (\citealt{2000ApJ...535..632M,2011NewAR..55..166F}). \\\
%
%
%
Early on, their high X-ray luminosities suggested that the mass of the compact object was higher than the average masses of the BHs observed in our Galaxy if standard accretion occurs. It was proposed \citep{1999ApJ...519...89C} that they could host Intemediate Mass Black Holes (IMBHs) with masses of $\sim 10^{3} - 10^{5} $ \msol . \\
However, further investigations revealed also that their X-ray spectral properties are unusual if compared to those observed in Galactic X-ray Binaries (XRBs) \citep{2001ApJ...554.1282M} opening the possibility to explain ULXs in terms of alternative accretion disc scenarios, without invoking IMBHs. It was proposed that super-Eddington accretion discs, as the slim discs models, around stellar-mass or slightly more massive BHs,  could reproduce the spectra of ULXs \citep{2003ApJ...597..780E}.\\
A significant effort was devoted in recent years to understand the nature of ULXs. With the advent of more sensitive X-ray telescopes (\textit{Chandra, XMM-Newton, NuSTAR}) the spectral properties of ULXs could be analyzed in greater detail.
\cite{2007Ap&SS.311..203R}, \cite{2009MNRAS.397.1836G} and \cite{2013MNRAS.435.1758S} identified the so-called \textit{Ultra luminous state}, which is characterized by the presence of some important features in ULXs spectra. This led to the conclusion that ULXs should be stellar-mass or massive (up to 100 \msol) BHs accreting with marginally or highly super-Eddington rates from massive donors.
In fact \cite{2008ApJS..174..223B}, \cite{2009MNRAS.395L..71M} and \cite{2009MNRAS.400..677Z} explored the massive BH scenario, finding that low-metallicity environments could lead to the formation of BHs in the range of 30-80 \msol and that marginal super-Eddington accretion onto these BHs could explain the observed properties of ULXs.\\
Understanding the nature of ULXs became more and more challenging after the discovery of the first pulsating ULX \citep{2014Natur.514..202B}, which unequivocally showed that ULXs can also be accreting neutron stars. This discovery has been confirmed by the identification of other five pulsar ULXs \citep{2017Sci...355..817I,2017MNRAS.466L..48I,2018MNRAS.476L..45C,2019arXiv190604791R,2019MNRAS.488L..35S}  whose emission properties can only been explained in terms of super-Eddington accretion \citep{2017MNRAS.467.1202M}.\\
ULXs appear to be the sources where super-Eddington accretion reveals itself in a variety of phenomenological modes. A strong evidence of such an extreme regime is also the presence of outflows, inferred from the detection of radio, optical and X-ray giant nebulae (\citealt{2002astro.ph..2488P,2005astro.ph.12552M,2020NatAs...4..147B}), broad optical emission lines \citep{2015NatPh..11..551F} and blue-shifted X-ray absorption lines (e.g. \citealt{2016Natur.533...64P,2017MNRAS.468.2865P}).\\
The main goal of this work is to investigate the nature of ULXs using their multiwavelength emission properties along with binary evolution in the context of super-Eddington accretion.\\ In section 2 we present the evolutionary tracks of simulated ULX binaries with BHs and their emission properties. The third section is devoted to the comparison of our model with some nearby and well studied ULXs. Finally, in the last section we explore the possibility to extend our model to the case of PULXs.

\section{Evolutionary tracks of ULXs and their optical appearence}
In this section we compute the evolutionary tracks of binary ULXs and their optical emission throughout it. Following previous work \cite{2008ApJS..174..223B}, \cite{2008MNRAS.386..543P} (hereafter PZ1), \cite{2009MNRAS.395L..71M}, \cite{2009MNRAS.400..677Z} and \cite{2010MNRAS.403L..69P} (hereafter PZ2) , here we assume that these systems are High Mass X-ray Binaries, and that the compact object is a black hole which is accreting matter from a massive donor via Roche lobe overflow. We further asssume that the mass transfer process starts when the donor is on the Main Sequence (case A mass transfer, see \citealt{2006epbm.book.....E}).
\label{sec:tracks}
\subsection{MESA evolutionary tracks}
\label{ssec:mesa}
We used the stellar evolution code MESA (Modules for Experiments in Stellar Astrophysics, release 10108, \citealt{2011ApJS..192....3P,2013ApJS..208....4P,2015ApJS..220...15P}), to run a set of representative ULX binaries accreting via Roche Lobe Overflow.\\ 
We considered donors with masses in the range $ 5\-30 M_{\odot}$ and BH masses in the range $10\--100 M_{\odot}$.
In this work the donor metallicity is fixed at the solar value (Y = 0.28, Z= 0.02). However, we tested the validity of this approximation and evolved some binaries with sub-solar metallicity. The resulting tracks  (see App. \ref{app:metallicity}) show that this is a valid approximation for the scope of this work. We also underline that \cite{2020A&A...638A..55K} studied the effects of metallicity on binary evolution, and found that it plays an important role only for those systems which start the accretion phase when the donor is an evolved star (case B mass transfer). As we focus our analysis on systems that start mass transfer on the main sequence (case A mass transfer),  in the following we consider the assumption of solar metallicity acceptable.\\
We took into account the effects of the wind mass loss, which affects significantly the evolution of massive donors.  Among the wind schemes that can be implemented, we chose the method that in MESA is called 'Dutch wind scheme', which differentiates the wind mass loss prescription for the stars in the Main Sequence (MS) from that after the main sequence, following the work by \cite{2009A&A...497..255G}. For the MS we used the default input relation \citep{2001A&A...369..574V,2000A&A...362..295V}. For the evolved stars we adopted the de Jager relation \citep{1988A&AS...72..259D} that, as observed by \cite{2017A&A...603A.118R}, is useful when the main objective is to study mass loss along the evolution on the HR diagram. 
We also note that in the specific systems we made evolving, i.e. during RLOF, mass loss via accretion is orders of magnitude higher than that expelled through stellar wind (see also \citealt{2014ARA&A..52..487S} for a comprehensive review), so that we do not expect that a different wind prescription can considerably alter our results.\\
%
First, we run the \textit{star} module for each donor mass. Then,we set the start of the RLOF phase during Main Sequence (MS), when the central Hydrogen abundance (h1) is $ \simeq 0.5$ (case A mass transfer). We denote the radius of the donor at this stage with $R = R_{d,h1} $. The evolution is stopped at the onset of helium burning in the core.\\ We set the initial period for the binaries 
combining Kepler's law:
\begin{equation}\label{eq:kepler}
   P \approx 2 \pi G^{-1/2} a_{1}^{3/2}(M_{d}+M_{a})^{-1/2}
\end{equation}
with the Eggleton approximation. Under the hypothesis that $R_{d,h1} = R_{L}$ (\citealt{1983ApJ...268..368E}):
\begin{equation}\label{eq:EggleR_l}
  a_{1}   \approx     R_{d,h1}  \frac{0.6q^{2/3} +\ln(1+q^{1/3})}{0.49q^{2/3}}
\end{equation} 
where $R_{L}$ is the Roche Lobe Radius of the donor, \textit{$a_{1}$} is the initial orbital separation, $M_{d}$ and $M_{a}$ are the donor and accretor mass, and $q = {M_{d}/M_{a}}$. 
The initial orbital periods turn out to be in the range 1-2 days.
The RLOF is calculated with the MESA implicit scheme 'Roche Lobe'. We set the minimum mass transfer rate at $\dot{M}_{min} = 1.0 \times 10^{-12} M_{\odot}/yr$.\\
The timestep is chosen in such a way to have an adequate time resolution without excessively increasing the computational time (typically about $\sim 4 \-- 5$ hours).
As the binary evolution proceeds faster after the Terminal Age Main Sequence (TAMS), the time resolution for the evolution of the donor during the post-MS stages is tighter than that adopted during the MS \footnote{setting '$varcontrol\_ms$' = 1d-4 and '$varcontrol\_postms$' = '$varcontrol\_casea$' = '$varcontrol\_caseb$' = 1d-5}.
We also set to zero the initial spin of the BH, which increases as accretion proceeds. 
\subsubsection{Orbital angular momentum through the outflow}
According to the model with an accretion disc plus an outflow adopted in \cite{mio_primo}, hereafter AZ, the evolution proceeds via non-conservative mass transfer and the accretion flow is not Eddington-limited.
When the mass transfer rate becomes super-critical, the mass ejected with the outflow effectively removes also orbital angular momentum. The total outflow rate depends on the fraction of radiative energy spent in launching the wind, $\epsilon _{w}$, and is given by (eq. 24 of \citealt{2007MNRAS.377.1187P}):
\begin{equation}\label{eq:mdot_wind}
	\dot{M}_{w} \approx a_{w} \dot{M}
\end{equation}
where $\dot{M}$ is the mass transfer rate and, following \citealt{2007MNRAS.377.1187P}, $ a_{w} = \epsilon _{w} (0.83-0.25 \epsilon _{w})$. We assume that half of the radiative energy accelerates the outflow ($\epsilon _{w} = 0.5$) and then $ a_{w} = 0.3525$.\\
As the outflow is launched from the inner disc where the radiation pressure is higher, we assume that matter leaves the system from the vicinity of the BH. The resulting orbital angular momentum loss is implemented in MESA following \cite{1997A&A...327..620S}: 
\begin{equation}\label{eq:jdot_ml}
\frac{\dot{J}_{ml}}{J _{orb}} = \frac{\alpha + \beta q^{2} + \delta \gamma (1+q)^{2}}{1+q} \frac{\dot{M} _{d}}{M_{d}}
\end{equation}
where $\alpha$ is the fraction of the mass lost from the donor, $\beta$ that expelled from the zones near the accretor, and $\delta$ that expelled from a circumbinary planar toroid with radius $r_{t} = \gamma ^{2} a$ (we refer to \citealt{1994inbi.conf..263V} and \citealt{2015ApJS..220...15P} for details). According to Eqs.~\ref{eq:jdot_ml} and ~\ref{eq:mdot_wind}, we then set  $\beta = a_{w} = 0.3525$, $\alpha = \delta = \gamma = 0$.

\subsubsection{Evolutionary tracks}
\begin{figure*}
	\includegraphics[width = \textwidth]{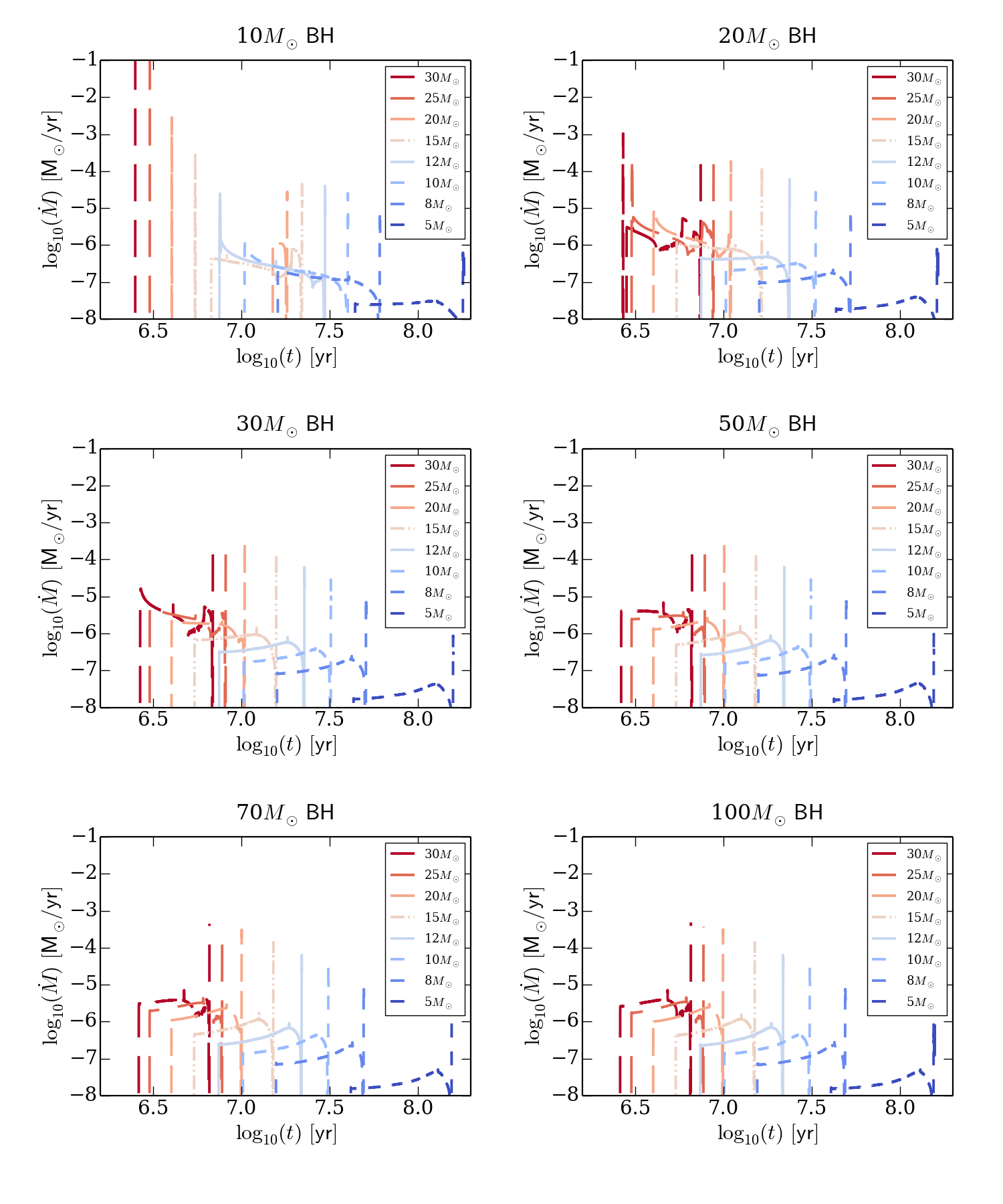}
	\caption{Evolution of the mass transfer rate for the binaries analyzed in this work. The BH mass is fixed in each panel. Short dashed lines refer to donors of 5 (dark blue), 8 (blue) and 10 (light blue) \msol , solid lines to donors of 12 (azure) \msol, dashed double-dotted lines to donors of 15 (salmone) \msol , and long dashed lines to more massive donors of 20 (orange), 25 (red) and 30 (dark red) \msol. }\label{fig:mbhconst_mdot}
\end{figure*}
The evolution of the mass transfer rate for the systems considered here is shown in Fig.~\ref{fig:mbhconst_mdot}. The tracks are grouped together according to the initial BH mass and color-coded according to the initial donor mass.
\begin{figure*}
	\includegraphics[width = \columnwidth]{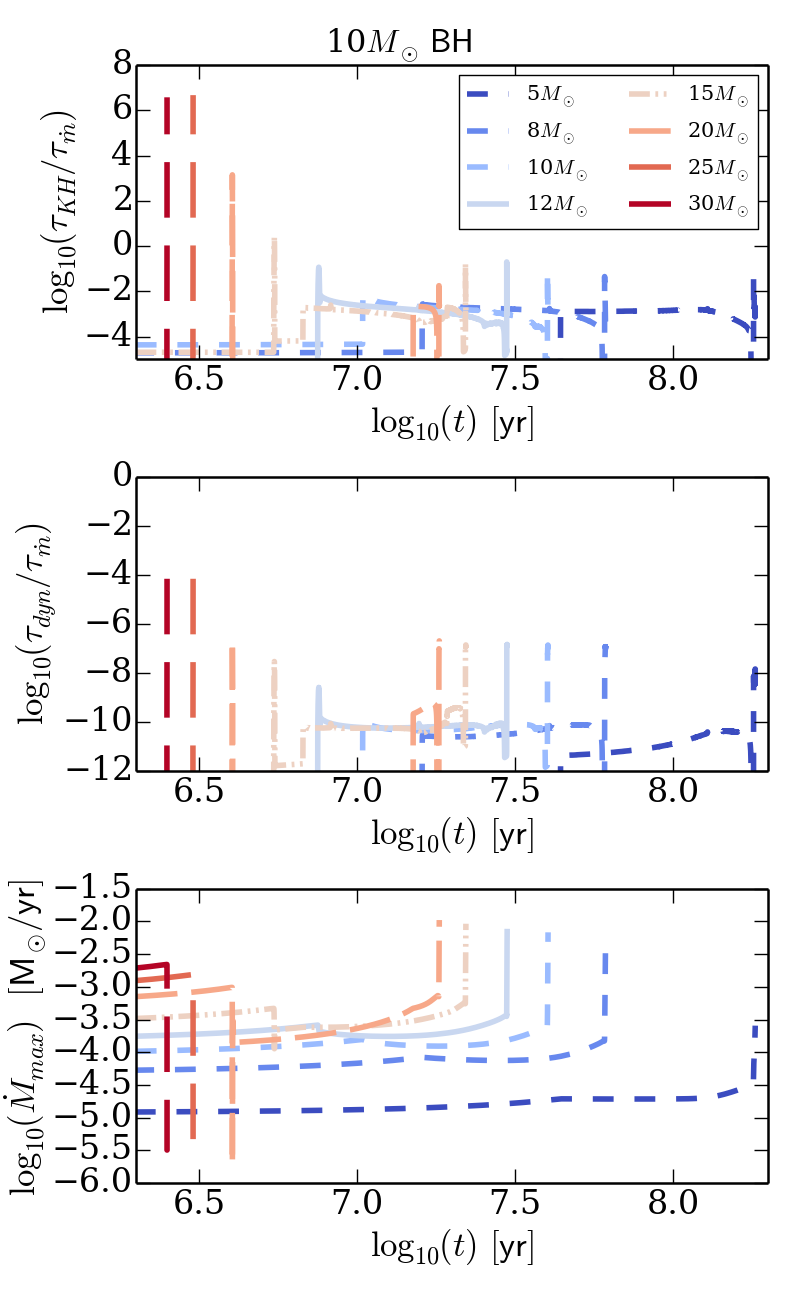}
	\includegraphics[width = \columnwidth]{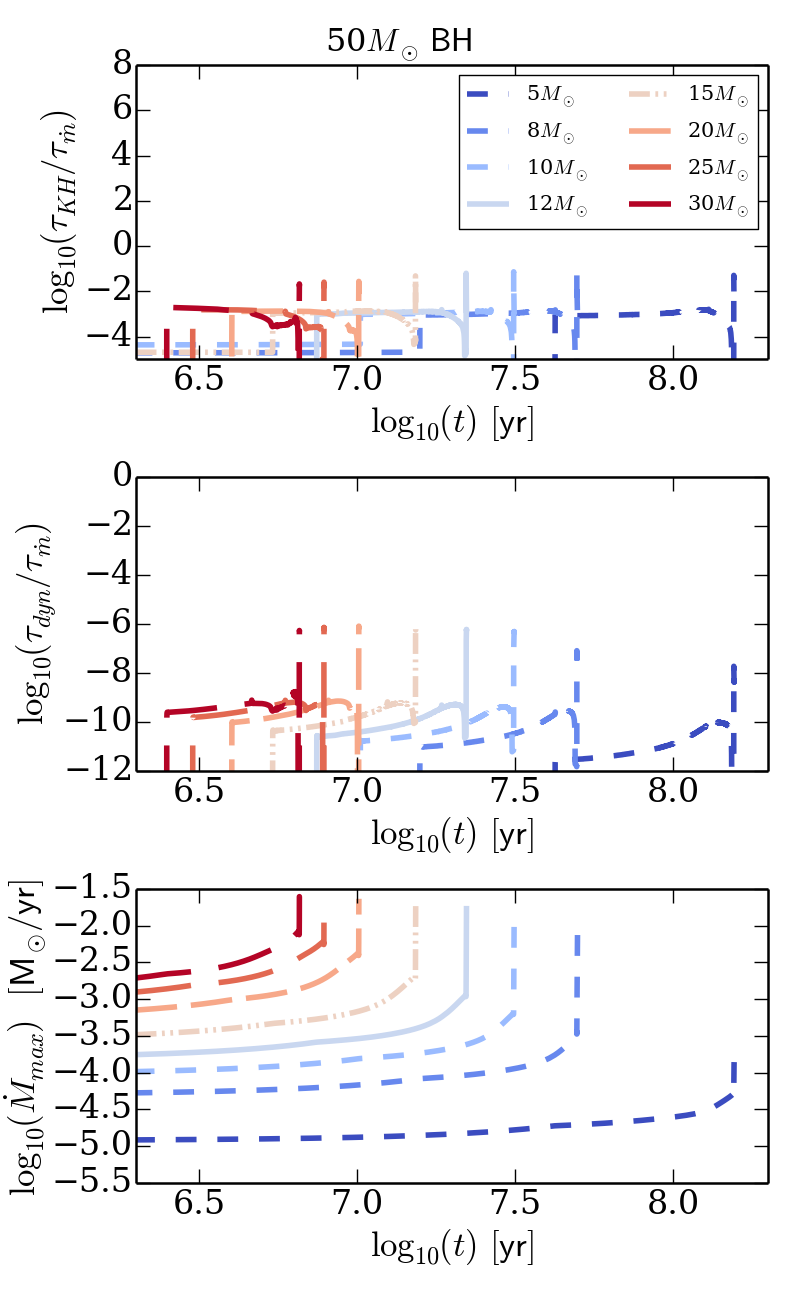}
	\caption{Evolution of the characteristic timescales for binaries accreting onto 10 (left) and 50 (right) \msol BHs. Upper panel: evolution of the ratio between the Kelvin-Helmotz ($\tau_{KH}$) timescale of the donor and the characteristic timescale for the mass transfer $\tau_{\dot{m}}$. Middle panel: evolution of the ratio between the dynamical timescale $\tau_{dyn}$ of the donor and the mass transfer timescale. Lower panel: evolution of the maximum value of mass transfer achievable without perturbing the hydrostatic equilibrium of the donor. }\label{fig:tscale}
\end{figure*}
\begin{figure*}
  \subfloat[]{%
    \includegraphics[width=0.9\columnwidth, height=0.78\columnwidth]{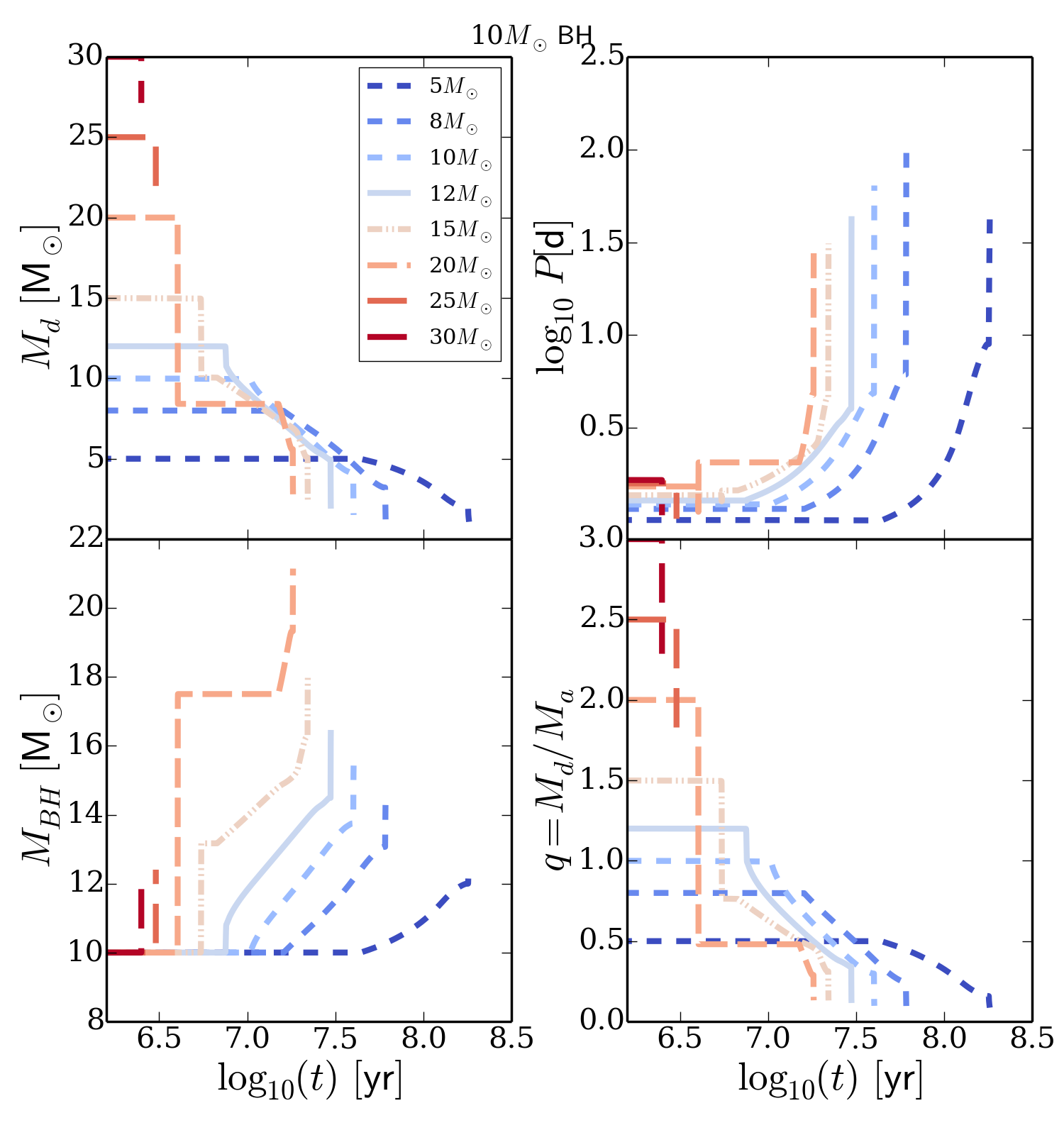}%
    \label{fig:10msolbh}%
  }
  \subfloat[]{%
    \includegraphics[width=0.9\columnwidth ,height=0.78\columnwidth]{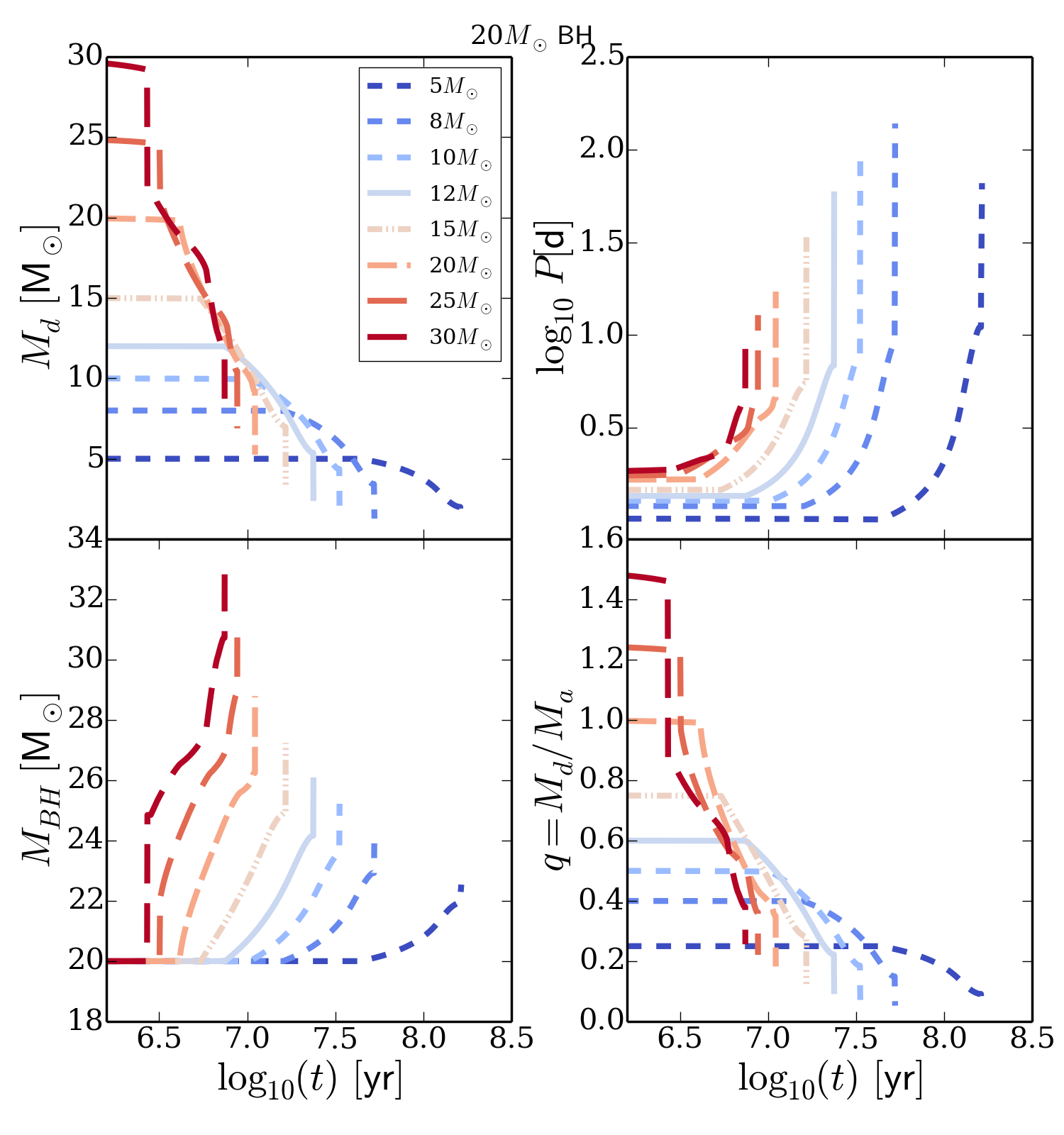}%
    \label{fig:20msolbh}%
  }\\
    \subfloat[]{%
    \includegraphics[width=0.9\columnwidth, height=0.78\columnwidth]{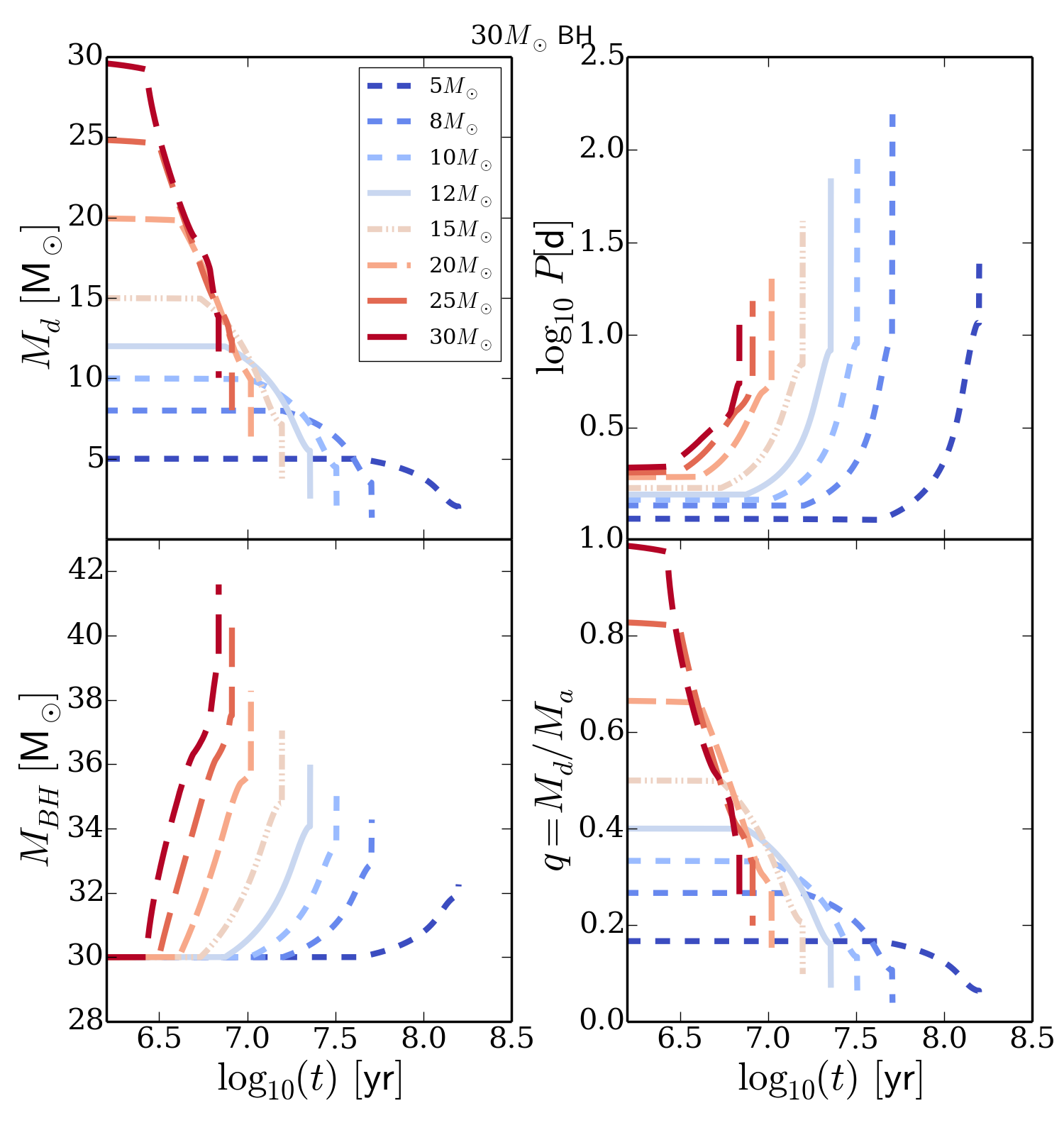}%
    \label{fig:30msolbh}%
  }
  \subfloat[]{%
    \includegraphics[width=0.9\columnwidth, height=0.78\columnwidth]{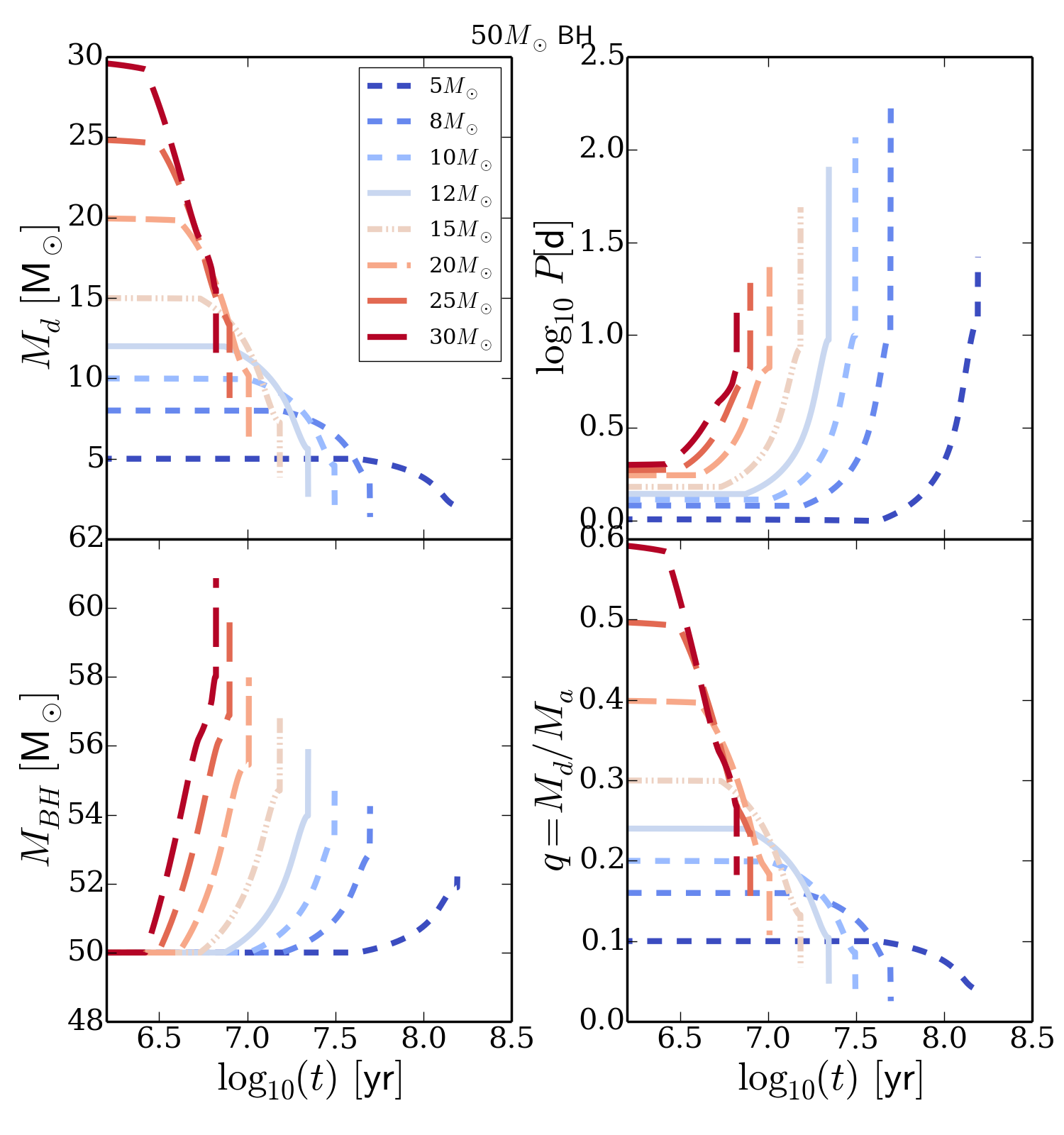}%
    \label{fig:50msolbh}%
  }\\
     \subfloat[]{%
    \includegraphics[width=0.9\columnwidth, height=0.78\columnwidth]{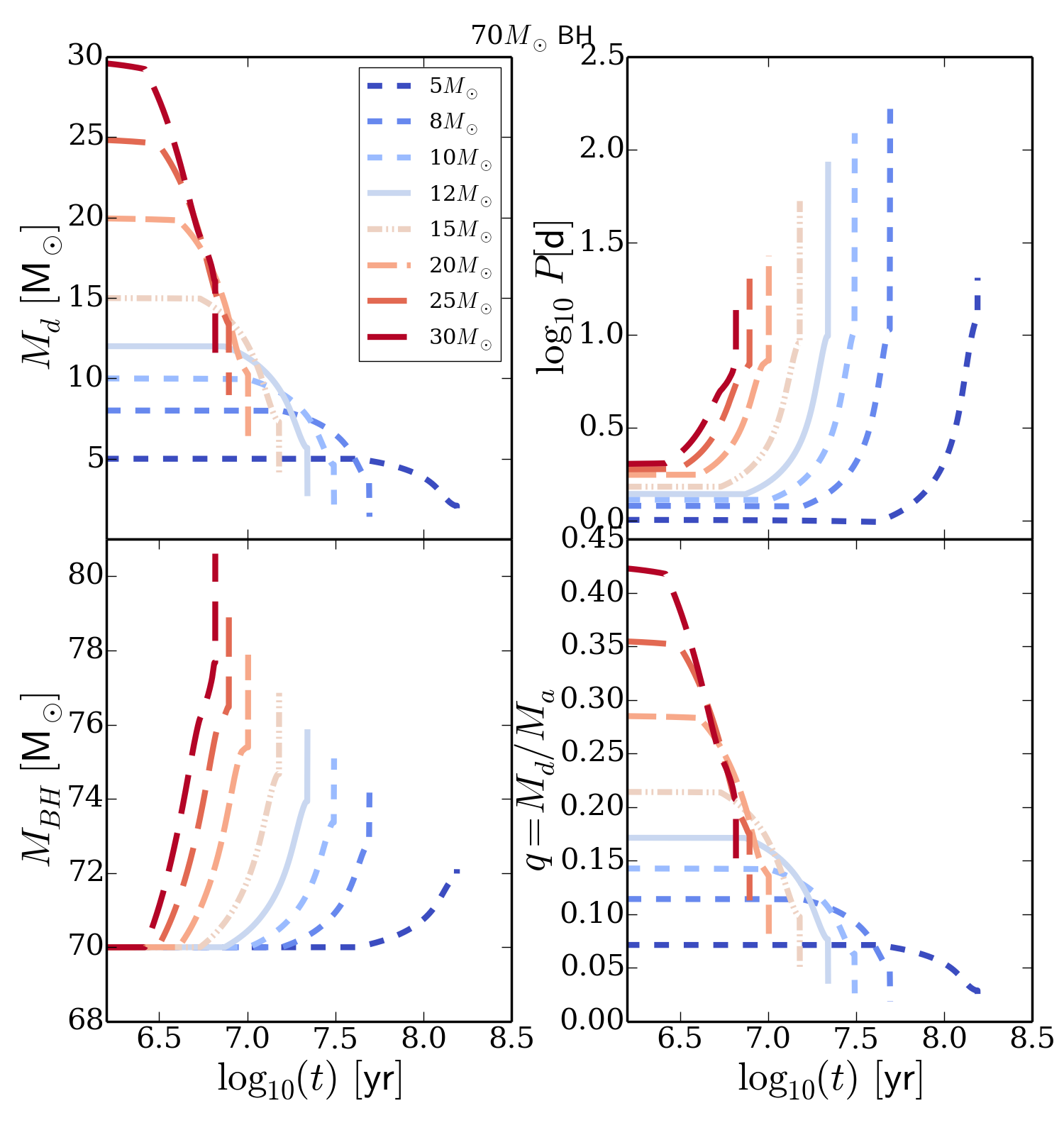}%
    \label{fig:70msolbh}%
  }
  \subfloat[]{%
    \includegraphics[width=0.9\columnwidth, height=0.78\columnwidth]{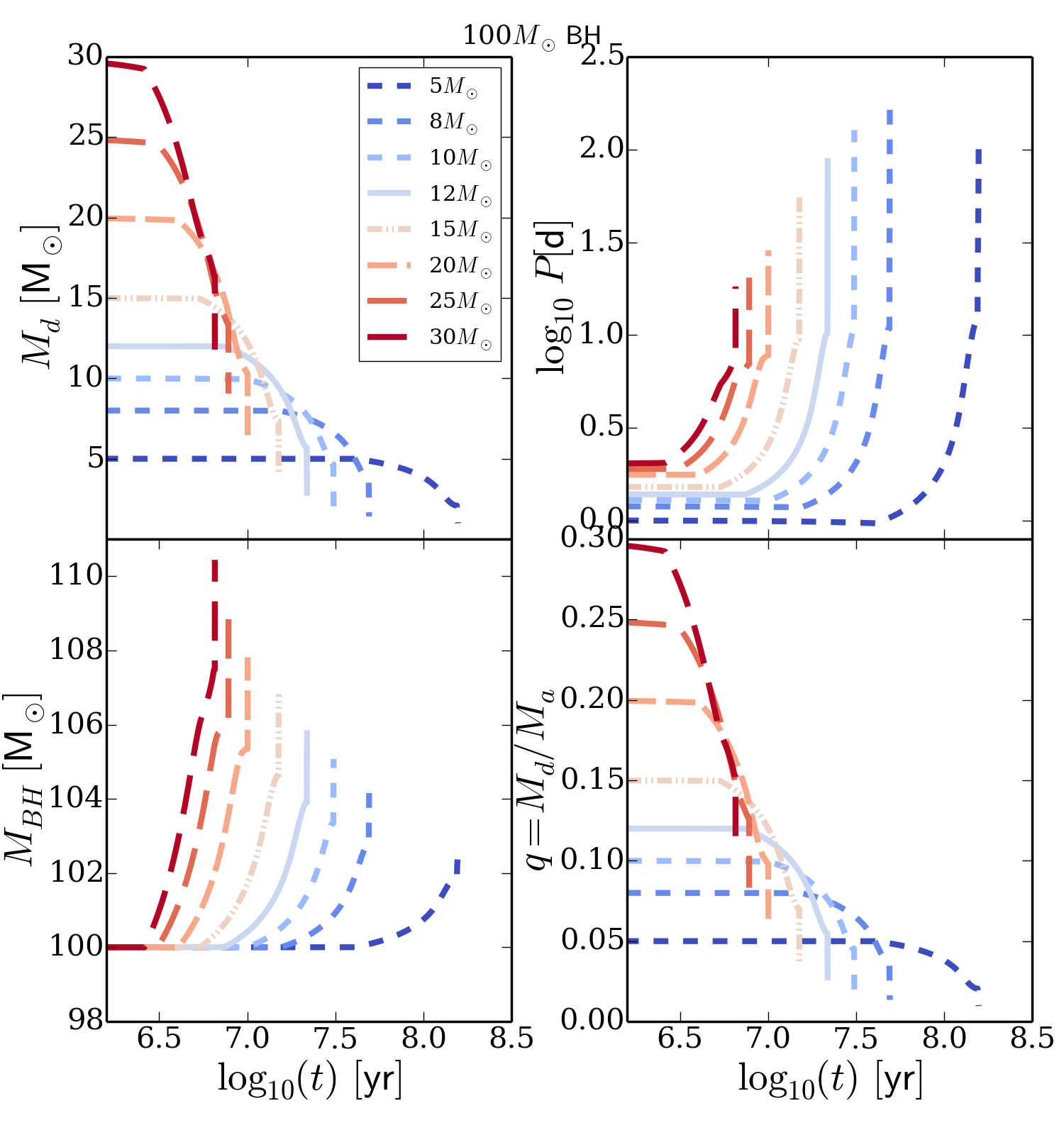}%
    \label{fig:100msolbh}%
  } \caption{\small{Evolution of the parameters for the binary systems evolved with MESA: donor mass
(upper left), orbital period (upper right), BH mass (lower left), and q ratio (lower right). Systems are grouped according to the BH mass and color-coded according to the donor mass as in Figs. \ref{fig:mbhconst_mdot} and \ref{fig:tscale}}} 
 \label{fig:parameters}
\end{figure*}
%
We describe the evolution of the tracks taking as reference the systems accreting onto a $10 M_{\odot}$ BH. They span a wide range in the mass ratio q, which is an important parameter driving the evolution of accreting binaries (see \citealt{2011epbm.book.....E}). We distinguish three different evolutionary paths.\\
For very massive donors of 25 and 30 \msol (q = 2.5 and q = 3 respectively), the accretion phase starts abruptly with an instability, reaching high rates (up to $10^{-1} \-- 1$ $ \text{\msol} /yr$, see top left panel of Fig.~\ref{fig:mbhconst_mdot}). The timescale for mass transfer ($\tau_{\dot{m}} = {M}/\dot{M}$) is faster than the thermal timescale, as shown in the top left panel of Fig.~\ref{fig:tscale}: $\tau _{KH}$\footnote{The Kelvin-Helmholtz time scale is the time scale on which a star reacts when there is disequilibrium between energy loss and energy production. It is defined as the ratio between the thermal energy content of the star $E_{th}$ and the luminosity L: $\tau_{KH} = \frac{E_{th}}{L} \approx \frac{GM^{2}}{2RL} \approx 1.5 \times 10^{7} \big( \frac{M}{M_{\odot}} \big)^{2} \frac{R_{\odot}}{R} \frac{L_{\odot}}{L}$ yr} is up to six orders of magnitude larger than $\tau_{\dot{m}}$.  Moreover, as the donor loses mass, it is driven rapidly towards instability, being the mass transfer rate higher than the maximum value permitted by the condition of hydrostatic equilibrium (see \citealt{Pols1} and the second and third left panels of Fig.~\ref{fig:tscale}). This effect is likely to lead the system towards a common envelope (CE) phase. This particular phase of binary evolution needs a dedicated modelling with physical assumptions different from those adopted for RLOF accretion, which is the relevant mechanism considered here for persistent ULXs. For this reason we stop the evolution of these systems at this stage, and postpone the inclusion of the CE phase in our model to a future paper.\\
Donors with initial mass of 15 and 20 \msol (q = 1.5 and q = 2 respectively) go through the first, thermal unstable mass transfer episode but do not become dynamically unstable: $\tau_{\dot{m}}$ is shorter than $\tau_{KH}$ (top left panel of Fig.~\ref{fig:tscale}) but the ratio between the dynamical timescale and the mass transfer timescale decreases driving the system towards stability (middle left panel of Fig.~\ref{fig:tscale}). As they lose mass, and the BH grows, the mass ratio decreases, the system detaches and the donor restores the thermal equilibrium. Thereafter, mass transfer is stable and proceeds on the nuclear timescale. We emphasize the extreme properties of the track of the 20 \msol donor: after having lost more than $50 \% $ of its mass during the first fast episode, the system widens and detaches. The donor restart RLOF at about TAMS.\\
For lower donor masses with $q \leq 1$ the evolution does not go through the first unstable phase of mass transfer, but it proceeds smoothly up to the TAMS, being the mass transfer timescale longer than both the Kelvin-Helmholtz and dynamical timescales.
Accretion during the giant phase proceeds with high accretion rates, but the donors maintain both hydrostatic and thermal equilibrium.\\
The evolution of systems accreting onto more massive BHs proceeds in a similar way but the initial accretion episode is less extreme for the more massive donors. For the 20\msol BH, the 25 \msol and 30 \msol donors can restore thermal and dynamical equilibrium, at variance with the 10 \msol case. For the 50 \msol BH the evolution is always stable (right panels of Fig.~\ref{fig:tscale}). Accretion is super-Eddington only during the giant phase.\\
Figs.~\ref{fig:10msolbh}-\ref{fig:100msolbh} show the evolution of a few chosen parameters of the binary systems: donor and BH mass (upper and lower left panels), orbital period and mass ratios (upper and lower right panels of each figure). The 5 \msol donor is almost totally stripped irrespectively of the BH mass (see dashed dark blue lines in the upper left panels).
Being its mass almost totally stripped, MESA could not find a convergent solution during the final phases and then its evolution was stopped before the onset of He burning in the nucleus. 
From Figs.~\ref{fig:10msolbh}-\ref{fig:100msolbh}  we see that irrespectively of the BH mass, the final orbital period is shorter for systems with more massive donors. 
For a given BH mass, the final orbital period for 8\msol donors can be one order of magnitude longer than that of a 30 \msol donor.
Finally, we emphasize that the accretion process changes dramatically the masses of the binary components. Donors lose more than 50 $\%$ of their initial mass, as shown in the upper left panels of Fig.~Figs.~\ref{fig:10msolbh}-\ref{fig:100msolbh} .
BHs acquire $\sim 2$ \msol from 5 \msol donors (dark blue lines in Figs.~\ref{fig:10msolbh}-\ref{fig:100msolbh} ) and $\sim 10$ \msol from 30 \msol donors (dark red lines in Figs.~\ref{fig:20msolbh}-\ref{fig:100msolbh} ). 
\subsection{Optical emission of ULXs}
\label{sec:mesa_tracks_cmd}
\subsubsection{Effects of super-Eddington accretion on optical emission}
\begin{figure}
 \includegraphics[width = \columnwidth]{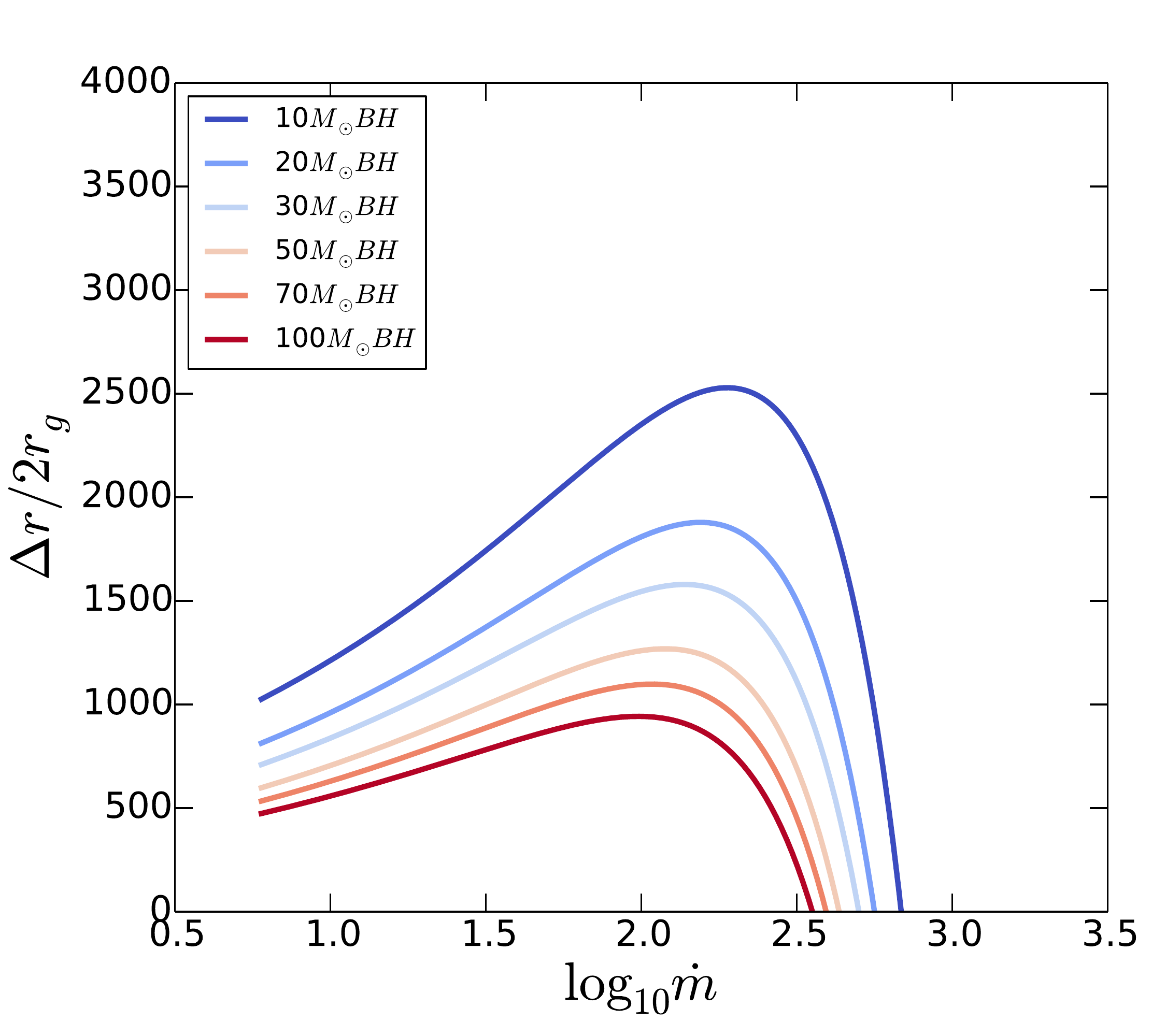}\caption{X-ray-UV irradiating region, $\Delta r = r_{x}-r_{ph}$, for the binary systems evolved with MESA. $r_{x}$ marks approximately the boundary of the region
where the gas temperature falls within the X-rays-ultraviolet (UV) energy band, and $r_{ph}$ is the outer radius of the optically thick photosphere of the outflow (details can be found in AZ). From top to bottom, the BH mass increases from 10 \msol to 100 \msol .}\label{fig:deltar_all}
\end{figure}
To determine the emission properties of accreting ULXs, we applied the model described in AZ to the evolutionary tracks of the systems reported in the previous section. 
The model computes the optical emission in the UBVRI Johnson and in the HST WFPC2, ACS and WFC3 photometric systems, that can be used to make a detailed comparison with the observed optical counterparts of ULXs. 
%
\begin{figure*}
	\includegraphics[width = \textwidth]{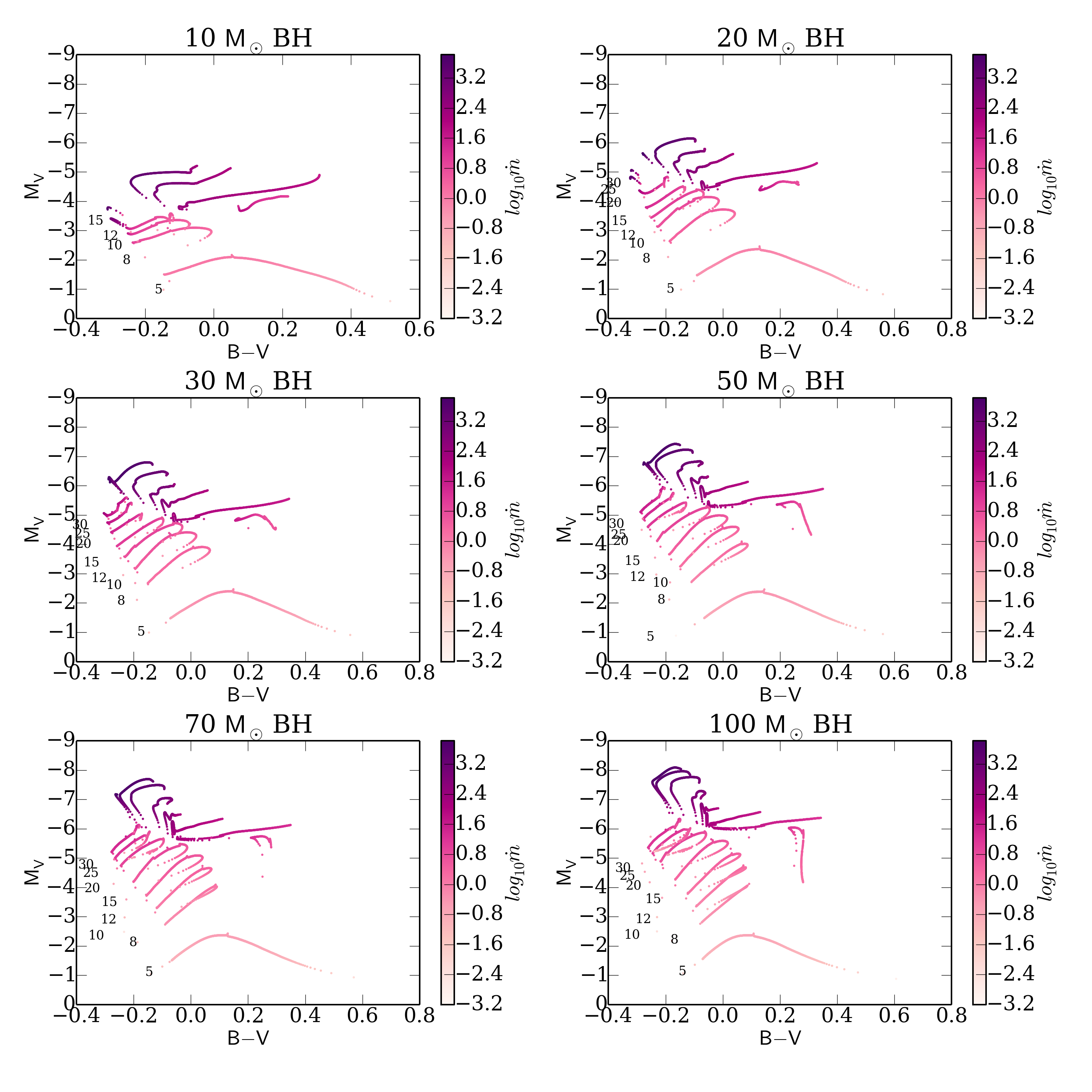}\caption{Evolutionary tracks on the color-magnitude diagram of the synthethic binary systems evolved with MESA: $M_{V}$ vs B-V color }\label{fig_CMD_MESA_bhfixed}
\end{figure*}
The optical emission of accreting binary systems is produced by different components, each contributing with a different weight to the total optical emitted flux, depending on the accretion regime and evolutionary stage. These components are: \begin{itemize}
	\item[-] the donor star, the properties of which are different from that of an isolated star with the same initial mass (see \citealt{2005MNRAS.362...79C}, PZ1, PZ2, \citealt{2011epbm.book.....E,Pols1} for details): during RLOF, its radius is constrained to be the Roche lobe radius and its temperature is determined by the condition of thermal equilibrium. Moreover, it can be heated up by the X-ray flux is produced by the disc;
	\item[-] the outer portion of the accretion disc that, for both sub-Eddington and super-Eddington accretion, is a standard Shakura $\&$ Sunyaev disc, and emits an optical flux proportional to the mass transfer rate. This region can reprocess and thermalize the X-rays from the innermost regions (see PZ1, PZ2, AZ and references therein);
	\item[-] the outflow, the  radial extension of which depends on the mass transfer rate. For extremely super-Eddington accretion, the outflow produces a large amount of optical radiation, which can significantly contribute to the total luminosity and, in the more extreme regimes, be the dominant component (see AZ).
\end{itemize}
As mentioned above, a strong contribution to the optical emission comes from the reprocessing of the X-ray-UV radiation produced by the inner regions of the disc and impinging in the outer disc and the donor star. If the disc is accreting at sub-Eddington rates, the flux intercepting the outer disc and the donor star makes the latter bluer than an isolated star with the same mass (see e.g. PZ1) and produces a bump in the red tail of the disc optical spectrum (see e.g. \citealt{1993PASJ...45..443S}).\\
Self-irradiation for super-Eddington accretion is more complex. In AZ we showed that the innermost regions of the disc do not irradiate the outer ones, because of the intervening outflow. The X-ray-UV irradiating flux is produced by a region located outside the outflow, the extension of which depends on the mass transfer. In addition, for extremely super-critical mass transfer rates, the region outside the photospheric radius of the outflow is too cold to produce a significant X-ray-UV flux: the disc-self-irradiation is therefore suppressed, but the intrinsic optical flux of the outer regions is very high.
\begin{figure*}
	\includegraphics[width = \textwidth]{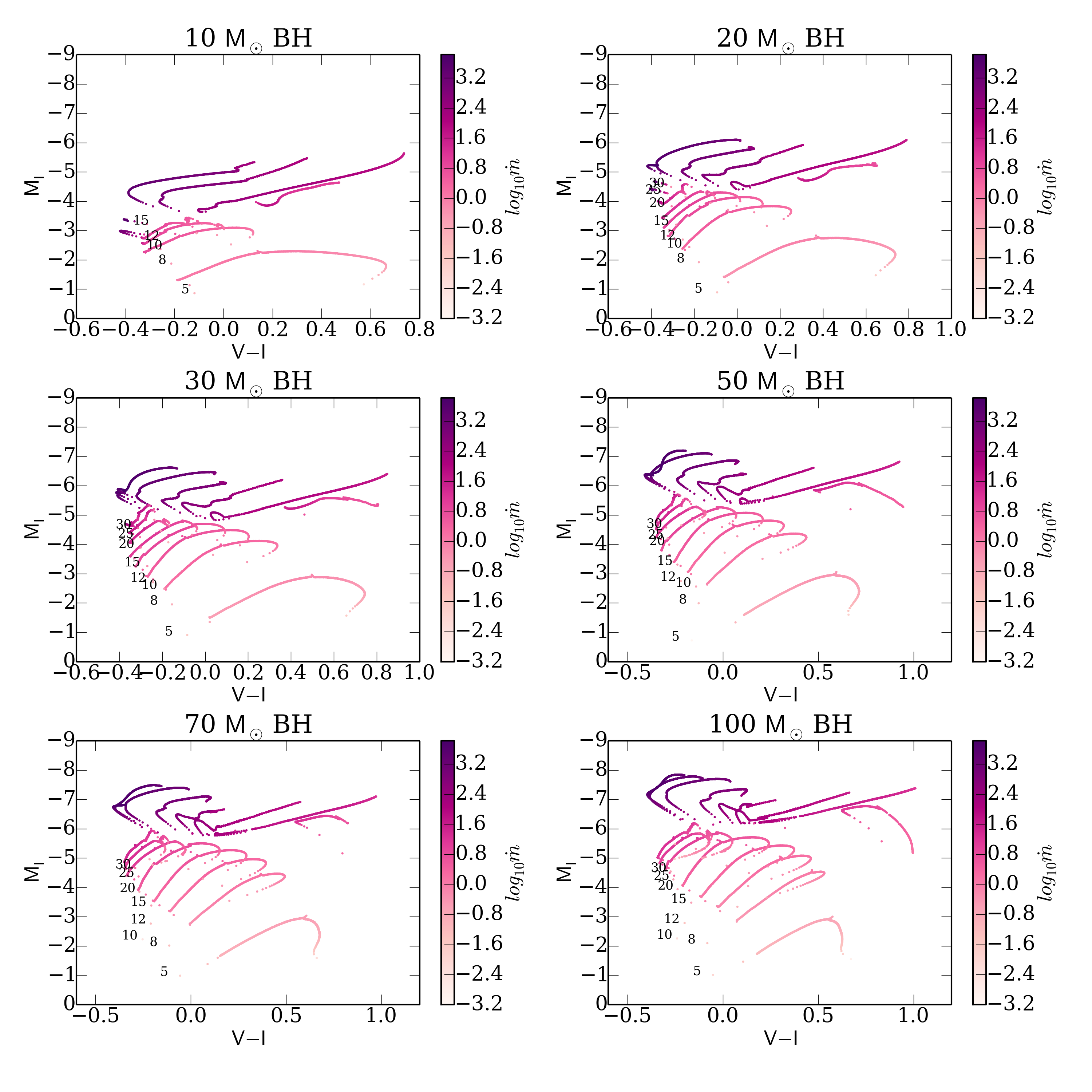}\caption{Evolutionary tracks on the color-magnitude diagram of the synthethic binary systems evolved with MESA: $M_{I}$ vs V-I color.}\label{fig:CMD_MESA_bhfixed_vmeni}
\end{figure*}
The size of the irradiating region in presence of an outflow is shown in Fig.~\ref{fig:deltar_all} for all the BH masses considered. 
The mass transfer is normalized to Eddington, and the irradiating region to the gravitational radius. From top to bottom, the BH mass increases from 10 to 100 \msol . As already discussed in AZ, for each BH mass, the irradiating region $\Delta R$ increases with the mass transfer rate, reaches a peak, and then decreases. The more massive is the BH, the smaller is the value of $\dot{m}$ at which the irradiating region reaches its maximum $(\dot{m} _{max, \Delta R})$. For a 100 \msol BH, $\log_{10} (\dot{m} _{max, \Delta R}) \sim 2.1$, while for a 10\msol BH  $\log_{10} (\dot{m} _{max, \Delta R}) \sim 2.4$, where $\dot{m}=\dot{M}/\dot{M}_{Edd}$.\\
\subsubsection{Evolutionary tracks on the CMD}
We computed the tracks on the Color-Magnitude Diagram (CMDs) of the systems described in the previous section from the onset of RLOF to the end of their evolution, which occurs while the donor ascends the giant branch. The duration of the tracks depends considerably on the donor mass: the more massive is the donor, the shorter is the evolutionary path, irrespectively of the BH mass. Systems with 30\msol donor last about $10^{6.5}$ yr, while systems with a 5\msol last about $10^{8.2}$ yr.
The evolutionary tracks on the CMD calculated with our code (AZ) and the MESA tracks are shown in Fig.~\ref{fig_CMD_MESA_bhfixed} ($B-V$ color and $M_{V}$ magnitude) and in Fig.~\ref{fig:CMD_MESA_bhfixed_vmeni} ($V-I$ color and $M_{I}$ magnitude).
They are ordered for increasing BH mass, which is fixed in each panel. The color code represents the magnitude of the accretion rate.\\
The evolutionary tracks of ULXs occupy two regions on the CMD, depending on the evolutionary stage of the donor. When the donor is on the MS and accretion is sub-critical or marginally super-critical, the tracks are blue and their $M_{V}$ magnitude is limited to $\sim -6$. They occupy the bluer and comparatively fainter corner of the CMD, and are represented by the light-pink coloured lines in Fig.~\ref{fig_CMD_MESA_bhfixed}. We notice that they are similar in shape for different BH masses, the main difference being the maximum value of $M_{V}$: more massive BHs generate more luminous tracks.\\
When the donor ascends towards the giant branch, the properties of the evolutionary tracks are governed by the mass transfer rate, which is now super critical, as shown in Fig.~\ref{fig:mbhconst_mdot}. Moreover, the binary separation increases rapidly  producing a more extended accretion disc, which emits a huge optical flux enhanced by self-irradiation.
This evolution drives the tracks towards higher V band luminosities. The colors are initially blue because the mass transfer rate at the beginning of the giant phase is very high and the disc is hot. As the super-Eddington mass transfer continues, the systems widen and the accretion disc becomes more extended, driving the tracks towards redder colors. In addition, self-irradiation is essentially suppressed. However, after the peak, the mass transfer rate starts to decrease, irradiation starts to contribute again to the optical flux and for systems with more massive donors the evolution towards the red slows down. These super-critical evolutionary stages are reproduced with the dark pink lines in Fig.~\ref{fig_CMD_MESA_bhfixed}. We notice that the track of the 5 \msol donor differs from the others. We will describe this case and the behaviour of the tracks for the various BH masses in Appendix \ref{app:systems}.
\section{Multiwavelength SED of nearby ULXs}
In AZ we concluded that the emission of the disc plus outflow alone cannot fully reproduce the observed data, because the X-ray spectrum at the highest energies is too soft and cannot properly describe the ULXs spectra. In many ULXs observations show evidence of a spectral component that can be interpreted as an optically thick and cool corona covering the innermost regions of the disc (e.g. \citealt{2009MNRAS.397.1836G}, \citealt{2012MNRAS.420.1107P}, \citealt{2013MNRAS.435.1758S}). Thus, we added
the contribution of a comptonizing corona, which covers the innermost regions of the disc, to the emission model of AZ . 
\subsection{Comptonization from an optically thick corona}
We approximate the corona with a sphere of radius $r_{c}$ which extends up to the radius where the outflow becomes optically thick:
\begin{equation}\label{eq:rc}
r_{c} = r_{ph,in},
\end{equation}
where $r_{ph,in}$ is the inner photospheric radius of the outflow (see AZ). Following the observational evidence gathered from the X-ray spectral analysis (e.g. \citealt{2012MNRAS.420.1107P}, \citealt{2014MNRAS.439.3461P}), we assume that the Compton parameter in the corona is significantly larger than unity and that Comptonization is saturated. In these assumptions, the high energy tail of the specific intensity of the source can be approximated with a Wien distribution ~\citep{1986rpa..book.....R}:
\begin{equation}\label{eq:Wien}
I_{\nu}^{W} = \frac{2h \nu ^{3}}{c^{2}} e^{- \alpha} e^{-h \nu / kT_{c}} ,
\end{equation}
where $T_{c}$ is the electron temperature of the corona.
Since Compton scattering conserves the photon number, the value of the factor  $e^{- \alpha}$ is determined from the relation:
\begin{equation}\label{eq:cons_pjot}
N_{ph}^{W} = N_{ph}^{d} .
\end{equation}
Here, $N_{ph}^{d}$ is the number of photons per unit time emitted from the disc which, for an observer at distance D, takes the form:
\begin{equation}\label{ndisc}
 N_{ph}^{d} = \pi  \int_{0}^{\infty}   \int_{r_{in}} ^{r_{ph,in}}  \bigg( \frac{I_{\nu} ^{d}}{h \nu} \bigg) \frac{2rD^{2}}{(r^{2}+D^{2})^{2}} dr d \nu ,
\end{equation}
while
\begin{equation}
 N_{ph}^{W} = \pi \frac{r_{c}^{2}}{r_{c}^{2}+D^{2}} \int _{0} ^{\infty} \frac{I_{\nu} ^{W}}{h \nu} d\nu
\end{equation}
is the number of photons which is emitted per unit time from the corona and $r_{in}$ is the inner disc radius. In the following, we consider two reference electron temperatures for the corona: 1.2 and 1.5 keV. These temperatures sample the typical range of temperatures inferred from modelling the observed ULXs spectra \citep{2014MNRAS.439.3461P}.
\subsection{Model-data comparison}
This section is dedicated to the comparison of our model with the optical and X-ray data of a few nearby ULXs. We constrain the model parameters using the photometric data and X-ray spectra of the sources.
\begin{table}
\scalebox{0.8}{
\begin{tabular}{lllll}
\hline 
ULX & Instrument & Obs.ID & Date & Exp. (ks) \\ 
\hline 
NGC 4559 X-7 & \textit{XMM-Newton} & 0842340201  &  2019-06-16 & 74.3  \\ 
&\textit{NuSTAR} &30501004002 &2019-06-17 & 94.9 \\
\hline 
NGC 5204 X-1 & \textit{XMM-Newton} & 	0693851401& 2013-04-21 & 16.9 \\ 
\hline 
Hol II X-1 &  \textit{XMM-Newton} & 0200470101 & 2004-04-15 & 10.4  \\ 
& \textit{NuSTAR} & 30001031002  & 2013-09-09 & 31.4 \\ 
& \textit{NuSTAR} & 30001031003  & 	2013-09-09 & 79.4\\ 
& \textit{NuSTAR} & 30001031005  & 2013-09-17&111 \\ 
\hline 
NGC 5907 X-2 & \textit{XMM-Newton} & 0795712601	 & 2017-12-01 & 60.8 \\ 
\hline 
\end{tabular}} \caption{Log of the X-ray observations used in this work.}
\end{table}\label{tab:xdata}
With the addition of an optically thick Comptonizing corona, our model can reproduce the Multiwavelength emission of ULXs in an  acceptable way, and can be effectively used to constrain the properties of ULXs.
For each source analyzed in this work, we proceed according to the following steps:
\begin{itemize}
	\item[1.] \textbf{Comparison with the CMD diagram}: the first step is to search the synthetic evolutionary tracks that intersect the optical emission of a ULX on the CMD. We search for intersections, when possible, in more than two colors, which provides an additional constraint.
	\item[2.] \textbf{Age selection:} the comparison with the evolutionary tracks typically produces several intersections. Therefore we select the model using an additional constraint, which is the age of the ULX inferred from the population study on the host environment.
	\item[3.] \textbf{Search for the best fit of the multiwavelength SED:} 
finally, after having selected the intersections matching the age of th ULXs, we computed the synthetic optical-through-X-ray SED of the correspondent snapshot of the track considering two electron temperatures of the optically thick corona: $kT_{c} = 1.2$ and $1.5$ keV. The most representative SED is chosen applying a chi-square test to the data-model comparison. Although we remark that we are not adopting a data spectral fit, the chi-square test empowers the selection process. 
	
We applied our model to those ULXs that satisfy some specific criteria: a) the compact object is unknown, b) the optical counterpart has been univocally detected and analyzed with HST data, c) the ULXs have high X-ray luminosity. These sources are NGC 4559 X-7, NGC 5204 X-1, Holmberg II X-1 and NGC 5907 ULX-2. In addition, the age of the population to which NGC 4559 X-7 and Holmberg II X-1 belong is known, and we will use it as a further constraint to the model selection. 
\end{itemize}
\begin{table*}
\scalebox{0.9}{
\begin{tabular}{ccccccccc}
\hline
\multicolumn{9}{c}{ \textbf{NGC 4559 X-7}}\\
\hline
\multicolumn{9}{c}{Best fit model for systems with $M_{d,i} = 15 $ \msol and $M_{BH,i} = 50 $ \msol}\\
\hline
\rule[-2ex]{0pt}{1.0ex} k T$_{e} (\mathsf{keV})$  &t (Myr)  &$M_{d}(t)$ (\msol) & M$_{BH}$(t) (\msol)& $\dot{m}$ & P (d) & L$_{\text{tot}} $ (erg s$^{-1}$) &   $\chi ^{2}$ &  d.o.f.\\
\hline 
1.2 & 15.3 & 5.5 &55.8 &$\sim  1 \times 10^{3}$  & $\sim$  18 & $  \text{3.2}\times \text{10}^{\text{40}}$  & 72.2 & 40\\ 
\hline 
\multicolumn{9}{c}{ \textbf{NGC 5204 X-1}}\\
\hline
\multicolumn{9}{c}{ Best fit model for systems with $M_{d,i} = 30 $ \msol and $M_{BH,i} = 30 $ \msol}\\
\hline 
\rule[-2ex]{0pt}{1.0ex} $k T_{e} (\mathsf{keV})$  &t (Myr)  &$M_{d}(t)$ (\msol) & $M_{BH}(t)$ (\msol)&$\dot{m}$ & P (d) & $L_{tot}$ (erg s$^{-1}$) &   $\chi ^{2}$ &  d.o.f.\\
\hline 
1.2 & 6.8 & 13.9 & 39 & 6.4 & 5.5 &$  \text{8.20}\times \text{10}^{\text{39}}$  & 106.4 & 21\\ 
\hline
\multicolumn{9}{c}{\textbf{Holmberg II X1}}\\
\hline
\multicolumn{9}{c}{ Best fit model for systems with $M_{d,i} = 20 $ \msol and $M_{BH,i} = 50 $ \msol}\\
\hline 
\rule[-2ex]{0pt}{1.0ex} $k T_{e} (\mathsf{keV})$  &t (Myr)  &$M_{d}(t)$ (\msol) & $M_{BH}(t)$ (\msol)&$\dot{m}$ & P (d) & $L_{tot}$ (erg s$^{-1}$) &   $\chi ^{2}$ &  d.o.f.\\
\hline 
1.2 &$ < 10$  & 11.5 & 55 & $\sim 10.5 $ & 5.5 &$  \text{1.3}\times \text{10}^{\text{40}}$ &31.7 & 36\\ 
\hline 
\multicolumn{9}{c}{ \textbf{NGC 5907 ULX-2}}\\
\hline
\multicolumn{9}{c}{ Best fit model for systems with $M_{d,i} = 15 $ \msol and $M_{BH,i} =  30$ \msol}\\
\hline 
\rule[-2ex]{0pt}{1.0ex} $k T_{e} (\mathsf{keV})$  &t (Myr)  &$M_{d}(t)$ (\msol) & $M_{BH}(t)$ (\msol)&$\dot{m}$ & P (d) & $L_{tot} $ (erg s$^{-1}$) &   $\chi ^{2}$ &  d.o.f.\\
\hline 
1.2 & 15.7 & 5.4 & 35.9 & 1890  &$\sim 14$  &$  \text{2.1}\times \text{10}^{\text{40}}$ & 8.3& 10 \\ 
\hline   
\end{tabular}}\caption{Best fit models found minimizing the $\chi ^{2}$ for the sources analyzed in this work.}\label{tab:chi_tutte}
\end{table*}
\subsubsection{NGC 4559 X-7}
We follow \cite{2005MNRAS.356...12S} and \cite{2011ApJ...737...81T}  and assume a Galactic extinction $E(B-V) = 0.018$  and a distance of 10 Mpc \footnote{However, redshift-independent measurements suggest that the distance of NGC 4559 can be $\sim 7$ Mpc (see \cite{2018A&A...609A..37C} and ref. therein). We will discuss the implications of the distance estimation on our results later.} of the host galaxy. We consider the average of the magnitudes and colors taken from two measurements \citep{2011ApJ...737...81T}: F555W = $-7.05\pm0.08$ and F555W-F814W = $-0.11\pm0.13$. We take as reference age for this source the age of the parent population: $ \sim 20$ Myrs \citep{2005MNRAS.356...12S}. 
The position of the source on the CMD intersects the tracks of donors with initial mass of 15 \msol accreting onto a 50 and 70 \msol BH during the shell H-burning phase and tracks of evolved donors with initial mass of 12, 15 and 20 \msol accreting onto a 100 \msol BH. The age of all tracks of X-7 at intersection ($\sim 15 $ Myr) are in agreement with that estimated from its stellar environment ($ \sim 20$ Myrs, \citealt{2005MNRAS.356...12S}). We then fitted their multiwavelength SEDs to further constrain the models.
We consider the \textit{XMM-Newton} and \textit{NuSTAR} observations of June 2019 ( see Tab. \ref{tab:xdata}). Indeed, the intersection of the ULX photometric point with a certain evolutionary track covers hundreds or thousands of years and fixes the average mass transfer rate at the Lagrangian point. The short term fluctuations of the accretion rate and physical conditions in the system must be averaged out when performing this comparison. Therefore, the error bar of the X-ray spectrum encompasses the daily/monthly/yearly variability range of the source flux that is typically a factor of $\sim$2.
Table \ref{tab:chi_tutte} shows the best fitting evolutionary tracks based on the $\chi ^{2}$ statistics: X-7 is reproduced with a system having donor of actual mass in the range 5-6 \msol accreting onto  a massive BH with actual mass of $\sim 56$  \msol. Accretion onto a 70 and 100 \msol BH are excluded by the spectral fitting.
\begin{figure*}
\includegraphics[scale=0.3]{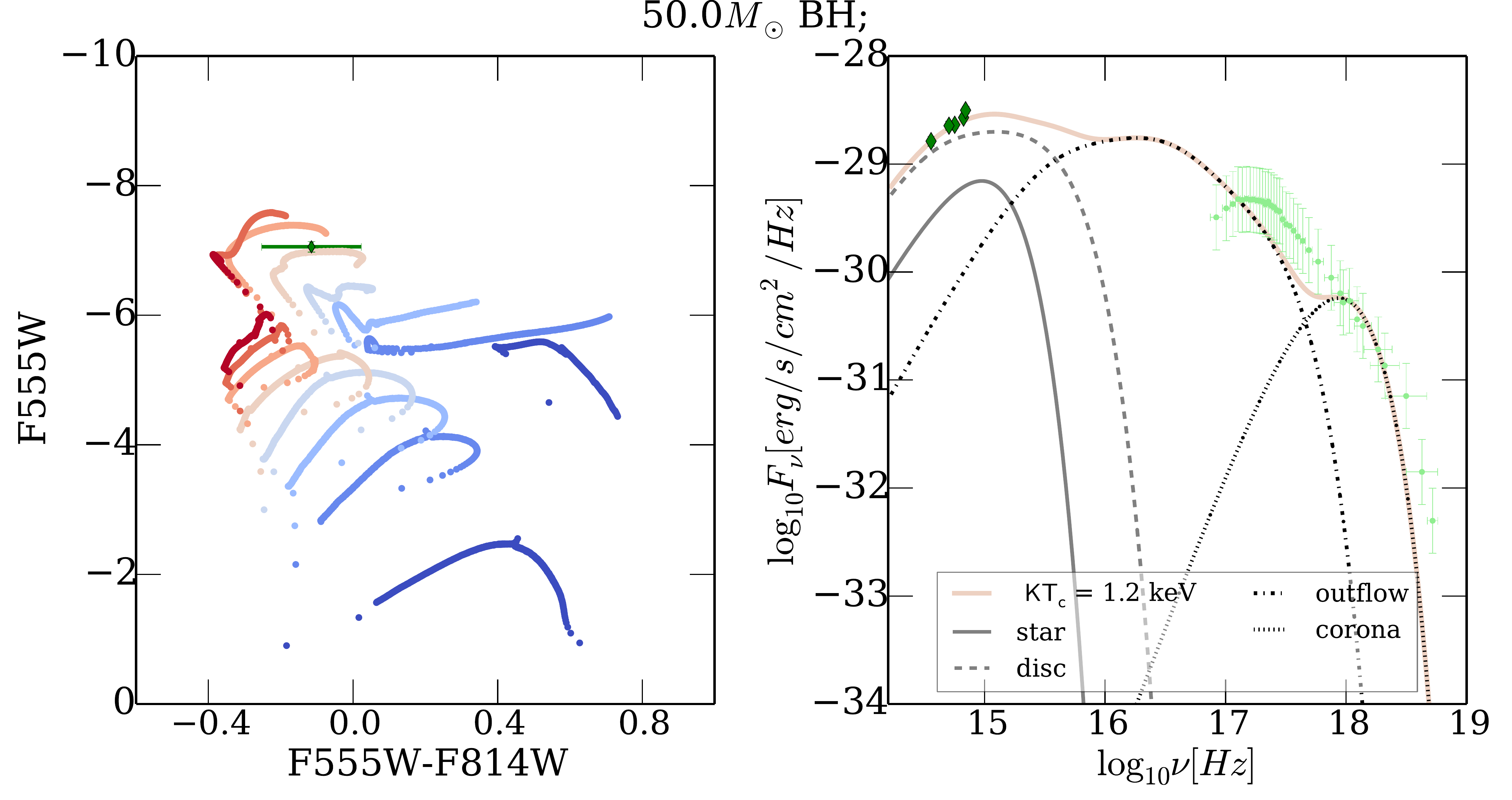}
\caption{Left: intersection of the optical counterpart of  NGC 4559 X-7 (green point) with the evolutionary tracks of a 50 \msol BH. Right: total SED for the best fit solution together with the single spectral components. The total SED has the same color of the evolutionary track (see left panel) to which it belongs. Green and lightgreen points indicates optical and X-ray data, respectively.}\label{fig:ngc4559_x7_sed}
\end{figure*}
%
The SED fit and the observed X-ray luminosity of X-7 ($\sim 2\times 10^{40} $ erg s$^{-1}$ \citealt{2005MNRAS.356...12S}) are better in agreement with a low corona temperature (Tab.~\ref{tab:chi_tutte}).
X-7 is well reproduced with a system accreting at largely super-critical rates onto a massive BH from donors which are ascending along the Giant Branch. Having lost most of their mass during the MS, are now less massive, with masses in the range 5-6 \msol . The orbital period of this systems is quite long ($ > 12$ days, see Tab.~\ref{tab:chi_tutte}) because the donor is evolved and has lost a large amount of mass.
Fig.~\ref{fig:ngc4559_x7_sed} shows that at these stages of the evolution, the optical emission is dominated by the outer accretion disc, which reaches a very high optical luminosity because of the high value of the mass transfer rate (dashed black line). At the same time, the very extended outflow dominates the UV/soft X-ray emission.
\subsubsection{NGC 5204 X-1}
NGC 5204 X-1 is located in the spiral galaxy NGC 5204 and has an average X-ray luminosity of $3\times 10^{39} $ erg s$^{-1}$. It is found to vary by $\sim 50 \%$ in 10 years \citep{2004ApJ...602..249L}. The source is thought to reside in a young stellar cluster with age $ < 10 $ Myr \citep{2002MNRAS.335L..67G}.
Among the data collected by \cite{2011ApJ...737...81T}, we will consider the averages of those taken in August 2008: F555W = $-5.646 \pm 0.055$ and F450W = $-5.848 \pm 0.027$ adopting a distance of 4.3 Mpc and galactic extinction $E(B \-- V )= 0.013$.
The optical counterpart of NGC 5204 X-1 is intersected by: the tracks of donors with initial mass of 20 and 25 \msol accreting onto a 20 \msol BH at super-critical rates while the donor is ascending the Giant Branch; the track of a 20 \msol donor accreting onto a 30 \msol BH in the first stages of its H-shell burning phase, or the track of a donor with initial mass of 30 \msol near the TAMS; the tracks of donors with initial mass 25 and 30 \msol accreting onto BHs of 50, 70 and 100 \msol during Main Sequence. These three cases differ in the evolutionary stages at which they occur: the more massive is the BH, the younger is the age of the system for the same initial donor mass, as we discussed in the previous section.\\
However, constraining the tracks with the age of the parent population does not help much, because the intersections are with tracks of very massive donors, whose evolution is very fast. We could rule out only the tracks of donors with initial mass of 25 \msol which accrete onto BHs of 20 and 30 \msol . We then calculated the SED for all the other intersections and searched for the models that best fit our data, using only the X-ray spectrum of \cite{2004ApJ...602..249L} (see Tab. \ref{tab:xdata}) and assuming an X-ray flux variability of $50\%$
.
%
\\
Running through all the SEDs selected according to photometry and age, with the two aforementioned electron temperatures of the corona, leads us to exclude the majority of the models. The system which better reproduces the optical and X-ray data is the one with initial donor and BH mass of 30 \msol , with actual donor mass of about 13 \msol and present BH mass of about 40 \msol which undergoes marginal super-Eddington accretion during TAMS. \\
However, the fit is formally not statistically acceptable and the bolometric luminosity is not well reproduced. The reason for which the fit is not satisfactory is related to the fact that the last two points of the X-ray spectrum are not well reproduced. The source is harder than what predicted by our model and seems to require an additional component to fit the X-ray data points at the highest energies, as noted by \cite{2015ApJ...808...64M}.
\begin{figure*}
	\includegraphics[scale = 0.3]{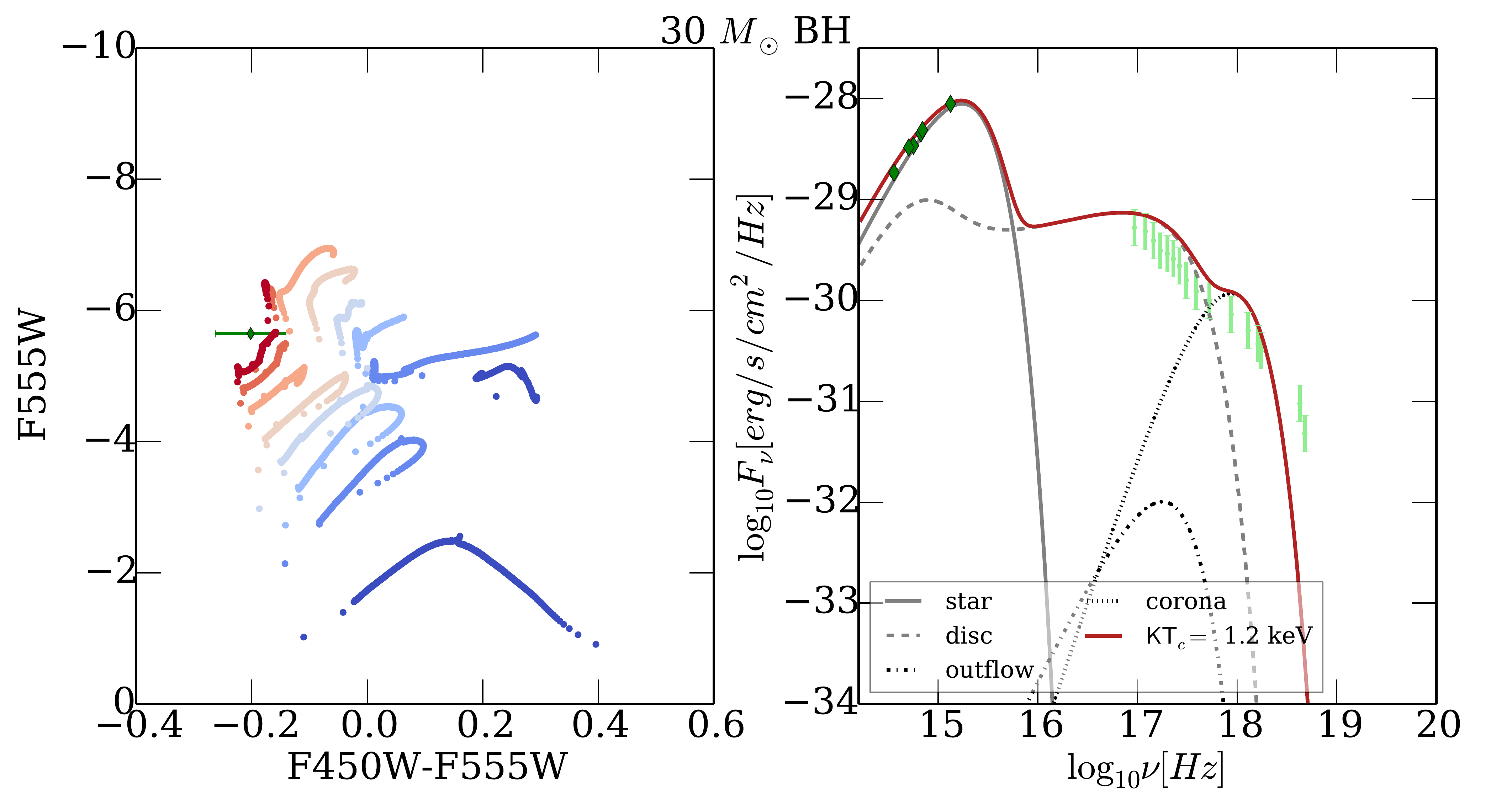}
\caption{Left: intersection of the optical counterpart of  NGC 5204 X-1 (green point) with the evolutionary tracks of a 30\msol BH. Right: total SED for the best fit solution together with the single spectral components. The total SED has the same color of the evolutionary track (see left panel) to which it belongs. Green and lightgreen points indicates optical and X-ray data, respectively.}\label{fig:5204_x1}
\end{figure*}
\subsubsection{Holmberg II X-1}
Holmberg II X-1 is located in the dwarf irregular galaxy, Holmberg II, which is in the M81 group of galaxies at a distance of 3.05 Mpc (see \citealt{2011ApJ...737...81T,2012ApJ...750..110T} and references therein). Its luminosity reaches $\sim 3\times 10^{40} $ erg s$^{-1}$  \citep{2010ApJ...724L.148G}. The source is surrounded by a photoionized optical nebula \citep{2004MNRAS.351L..83K} containing a point-like optical counterpart. A study published by \cite{2017MNRAS.467L...1E} shows that X-1 is escaping from a very young cluster aged $\sim 3.5\--4.5 $ Myrs \citep{2000ApJ...529..201S}, which will be the reference age for our study. We take the photometry from \cite{2012ApJ...750..110T}, that reported the average fluxes of non simultaneous observations in the F814W, F555W, F450W and F336W filters: $(2.71 \pm 0.54) \times 10^{-18}$, $(7.74 \pm 1.55) \times 10^{-18}$, $(1.34 \pm 0.27 ) \times 10^{-17}$ and $(2.70 \pm 0.54 ) \times 10^{-17}$  erg s$^{-1}$  cm$^{-2}$ \text{\AA}$ ^{-1}$, respectively. We search for intersections of the evolutionary tracks with data points on the CMD (calculated for different filters). Note that, because of the significant optical variability of the source, the error bar on the color is large and the intersections of the datapoint with the simulated evolutionary tracks are numerous.
Considering the very young age inferred by \cite{2000ApJ...529..201S}, the majority of the intersections are ruled out and we are left with systems accreting onto massive BHs (50 and 70 \msol) from donors with initial mass from 15 to 25 \msol. 
We then compared the SEDs obtained with our model with the observed optical through X-ray SED of Holmberg II X-1. The X-ray data are taken from \cite{2014MNRAS.439.3461P} (see Tab. \ref{tab:xdata}). Given the very young age, the majority of the best fitting SEDs belong to systems in the sub-Eddington accretion phase, which can reproduce well the softer part of the X-ray spectrum but have a significantly different Spectral Energy Distribution at the higher frequencies. 
The intersection that best fits the data is a system with a main sequence donor younger than 10 Myrs with initial mass of 20 \msol accreting onto a BH with initial mass of 50 \msol . The system is accreting marginally above Eddington ($\dot{m} \sim 10$). The actual mass of the donor is about 11 \msol and that of the BH about 55 \msol. The orbital period is short ($P \sim 5.5$ d), because during the long and stable MS RLOF phase, the mass ratio of the system does not change dramatically (see Tab.~\ref{tab:chi_tutte}).
\begin{figure*}
\includegraphics[scale = 0.3]{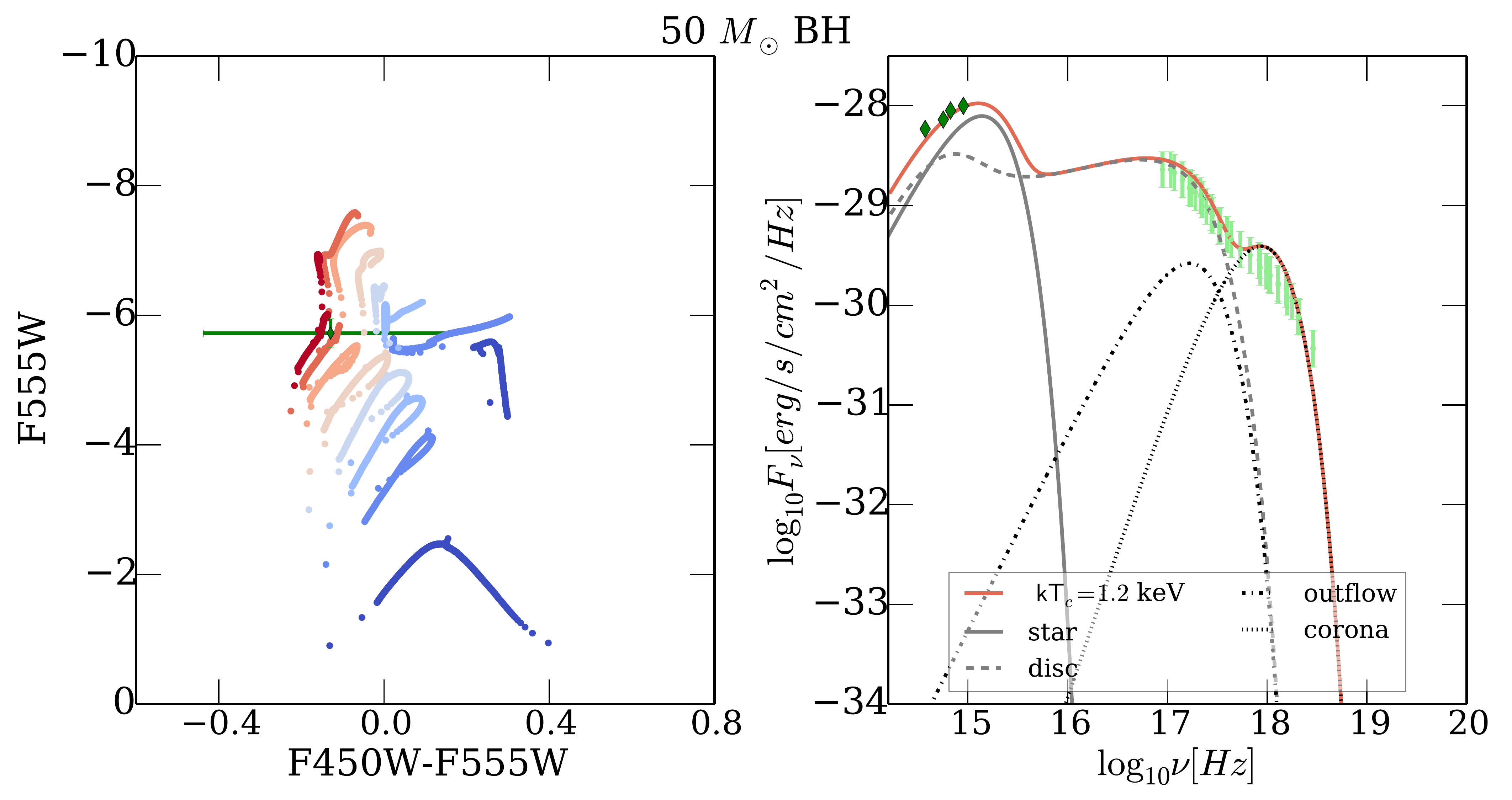}
 \caption{Left: intersection of the optical counterpart of  Holmberg II X-1 (green point) with the evolutionary tracks of a 50 \msol BH. Right: total SED for the best fit solution together with the single spectral components. The total SED has the same color of the evolutionary track (see left panel) to which it belongs. Green and lightgreen points indicates optical and X-ray data, respectively.}\label{fig:hol2_50sed}
\end{figure*}
Fig.~\ref{fig:hol2_50sed} shows that the SED that best fits the data is produced by a cool corona with an electron temperature of 1.2 keV. Moreover, being the accretion rate marginally super-critical, the outflow is not very extended and does not contribute much to the X-ray emission. Looking at the optical emission, it is dominated by the donor star but also the standard disc contributes in a significant way.

\subsubsection{NGC 5907 ULX2}
NGC 5907 ULX2 \citep{2018MNRAS.477L..90P} is located at a distance of 17.1 Mpc and it reached a peak X-ray luminosity of $6.4 \times 10^{39} $ erg s$^{-1}$. 
Being a recent discovery, it is not well studied, and we do not have much information on its variability during the active phase. However, we know that the source is transient because in a previous observation taken in 2012, Chandra did not detect it (with an upper limit of $1.5 \times 10^{38} $ erg s$^{-1}$ on the X-ray luminosity). Clearly, our model does not take into account the disc instabilities that may lead to a transient behaviour. However, we can use it to approximately describe its active phase. We used the optical photometry and X-ray data presented in \cite{2018MNRAS.477L..90P} (see see Tab. \ref{tab:xdata}). 
Some of the synthetic tracks intersect with the photometric point of NGC 5907 ULX-2 on the CMD computed using the F450W  magnitude and F450W-F814W color. The source cannot be reproduced by systems accreting onto a 10 or 20 \msol BH. On the basis of the photometry the system can be a BH of 30 \msol accreting from a donor which has an initial mass of 15 \msol , 50 \msol from a donor with initial mass of 15 \msol, 70 \msol from donors with initial mass of 12 and 20 \msol and 100 \msol from a donor with initial mass of 12 and 15 \msol .\\
For this source we did not find information on the age of the host environment. Therefore in the following we considered all the snapshots that intersect the photometric point.
Tab.~\ref{tab:chi_tutte} shows the result of the $\chi ^{2}$ test applied to the snapshots run for this source. 
The observed multiwavelength SED is better reproduced with a system composed of a donor star with mass of about 5.4 \msol which is accreting at super-Eddington rates onto a BH of about 36 \msol. The system initially consisted of a 15 \msol donor and a 30 \msol BH. The accretion phase takes place while the donor is ascending along the Giant Branch. The binary has an orbital period of $\sim 24 \, \text{d}$, therefore the disc is very extended.
\begin{figure*} 
 \includegraphics[scale = 0.3]{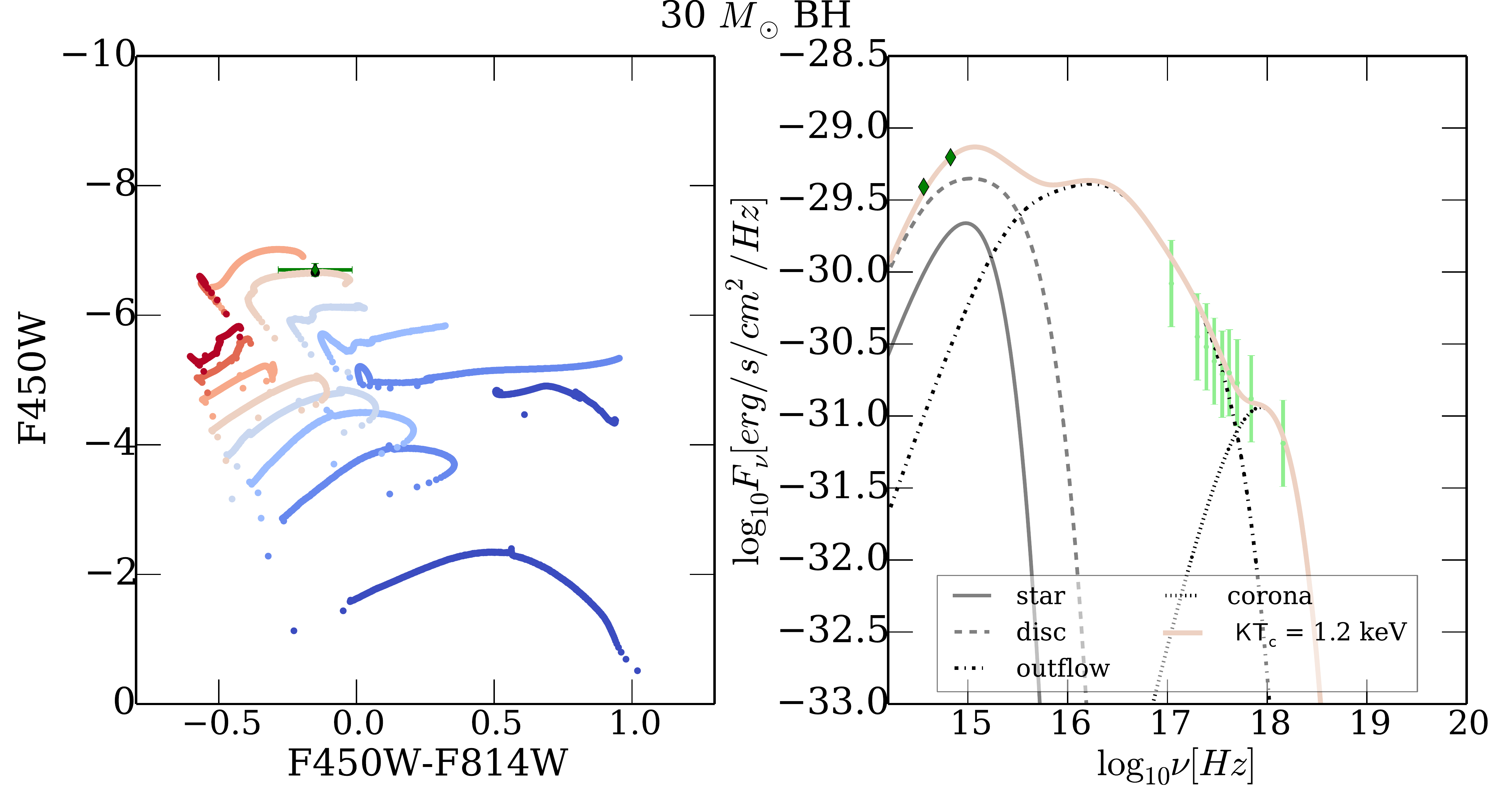}
\caption{Left: intersection of the optical counterpart of  NGC 5907 X-2 (green point) with the evolutionary tracks of a 30 \msol BH. Right: total SED for the best fit solution together with the single spectral components. The total SED has the same color of the evolutionary track (see left panel) to which it belongs. Green and lightgreen points indicates optical and X-ray data, respectively.}\label{fig:ngc5907ulx2sed}
\end{figure*}
In Fig.~\ref{fig:ngc5907ulx2sed} we show the  different components which contribute to the overall emission spectrum: the optical band is dominated by the outer accretion disc, which is very bright because of the very high mass transfer rate.


\section{Unveiling ULX Pulsars from Optical Data: an Exploratory Study}
Since the discovery of the first pulsating ULX in 2014 \citep{2014Natur.514..202B} five other sources have been found to be accreting pulsars 
\citep{2017Sci...355..817I,2017MNRAS.466L..48I,2018MNRAS.476L..45C,2019MNRAS.488L..35S,2019arXiv190604791R}.
Detecting pulsations likely requires a specific viewing angle and/or special physical conditions. If pulsars are not detected, it would be useful to have alternative approaches to distinguish between ULXs accreting onto a BH or a NS. 
In \cite{2017ApJ...836..113P} a large sample of bright ULXs have been studied with the aim to search for specific signatures in the X-ray spectra that could be linked to the nature of the compact object. Above 2 keV, the spectra of the known ULX pulsars are harder than those of other ULXs analyzed in their study. However, despite the effort, a clear spectral signature has not been found so far. 
Modelling the evolution and multiwavelength emission properties of ULX binary systems represents, in principle, an additional tool to assess the nature of the compact object and, at the same time, to understand the evolutionary paths that can lead to the formation of these systems (see PZ1, PZ2 and AZ). A detailed investigation of pulsars ULXs is beyond the scope of the present paper. Here we present only a preliminary study of the optical emission properties of ULX binary systems with a neutron star, that undergo a transient (quasi-steady) RLOF accretion phase similar to that of massive donors in BH binaries.
The purpose of the present investigation is simply to study the evolution of quite stable, low donor mass NS ULXs to get a grasp of their possible evolutionary paths and emission properties. High donor mass non-conservative systems are well beyond the present goals. Of course, the results presented here will not be relevant or applicable if all NS ULX binaries will definitely turn out to be system of the latter type.
\subsection{Optical Emission of Pulsar ULXs}
We used the evolutionary code MESA \citep{2011ApJS..192....3P,2015ApJS..220...15P} to evolve binaries with different masses accreting onto a neutron star. In this first exploratory study we did not consider the magnetic field of the neutron star, because we are only interested in the optical emission that originates in the outer region. Moreover, we simplify the problem considering that the mass transfer is conservative, since non-conservative mass transfer onto  NS from a massive donor is beyond the scope of this paper.  Here, we show how the optical emission evolves under the following assumptions: 
\begin{itemize}
	\item[-] a non-magnetized NS of 1.4 \msol;
	\item[-] conservative mass transfer;
	\item[-] case A mass transfer until the NS becomes very massive (we choose an upper limit of 2.5 \msol) or mass transfer becomes unstable;
	\item[-] donor stars with masses in the range 2-4 \msol . \end{itemize} This range of mass is broadly consistent with the donor of NGC 5907 ULX-1, which is thought to be an intermediate mass star in the range 2-6 \msol \citep{2017Sci...355..817I}; although the extinction in the direction of the source is very high \citep{Heida_2019_donorsULP}. In addition, this assumption is also approximately consistent with the non-detection of the counterpart of M82 X-2, if its donor belongs to the lower end of the range obtained by \citealt{2014Natur.514..202B}: $M_{d} \geq 5.2 $ \msol . \\
More massive donors cannot steadily be evolved with this model. Given the low mass of the neutron star, we found that systems with donors more massive than 4 \msol started abruptly dynamic unstable mass transfer.
We show the evolution of two representative systems with an intermediate donor mass and a quite different q ratio.
\begin{figure}
	\includegraphics[width=\columnwidth]{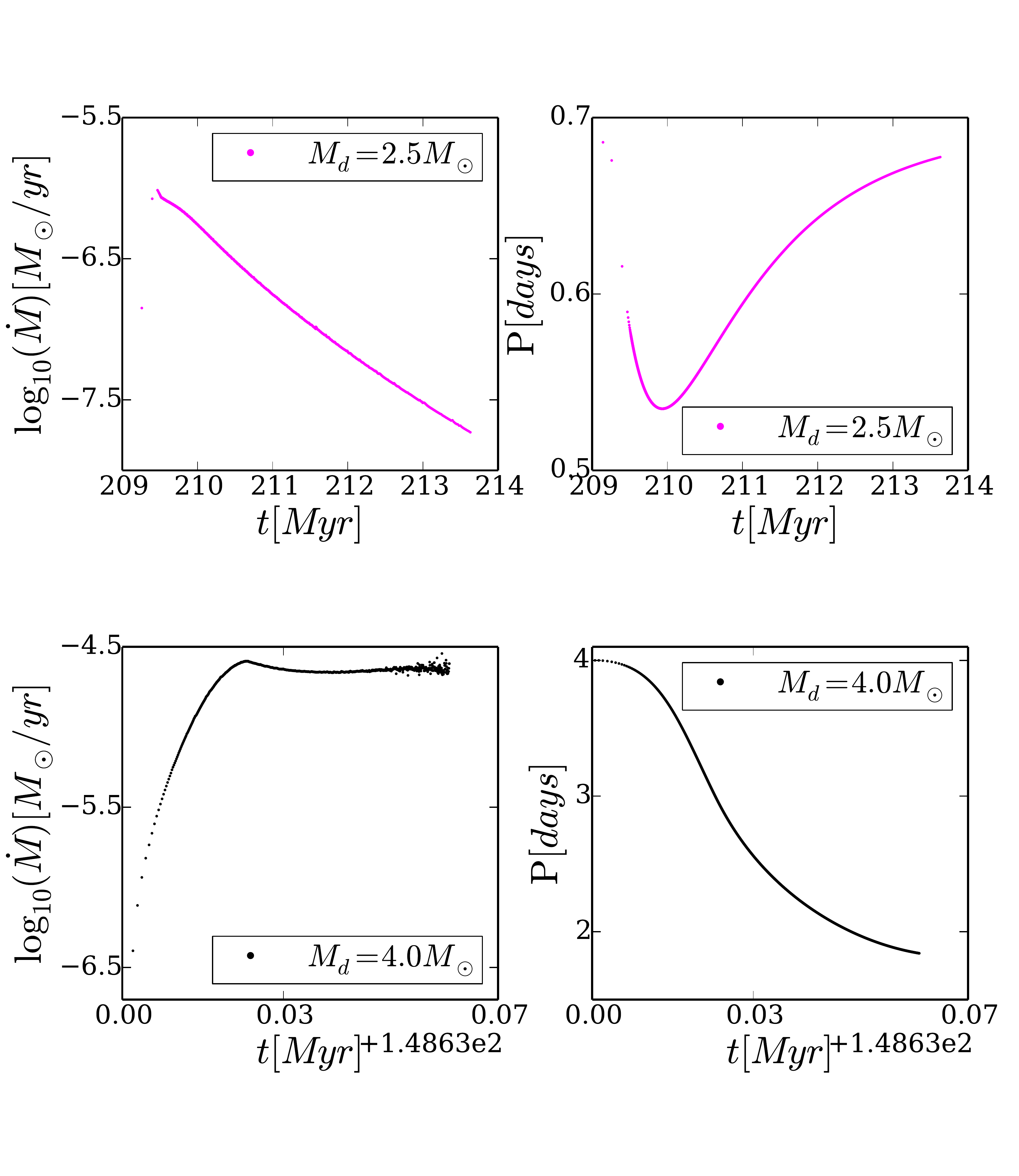}\caption{Upper panel: evolution of the mass transfer rate (left) and the orbital period (right) for a system with a 2.5 \msol donor and a 1.4 \msol neutron star. Lower panel: the same for a system with a 4.0 \msol donor. The mass transfer is super-critical at the onset of RLOF being the Eddington limit for a neutron star $\dot{M}_{Edd,NS} \approx 10^{-8}$ \msol /yrs
}\label{fig:pulx_ev}
\end{figure}
%
\subsubsection*{Systems with donor of 2.5 \msol}
When the RLOF phase of this system starts, accretion is super-Eddington, as it can be seen from the upper left panel of Fig.~\ref{fig:pulx_ev}.
The system initially shrinks (upper-right panel of Fig.~\ref{fig:pulx_ev}), then it starts to expand while the mass transfer decreases. After $\sim 4$ Myr from the onset of RLOF the neutron star mass reaches the critical value of 2.5 \msol and the evolution is stopped. 
\subsubsection*{Systems with donors of 4 \msol}
The system with a 4 \msol donor starts accretion with a higher mass transfer rate than the previous case.
During the short-lived RLOF phase ($\approx 7.0 \times 10^{4}$ yr) the system shrinks and the donor becomes unstable after having lost most of its mass.
%
\subsubsection*{Evolutionary tracks on the CMD}
\begin{figure}
\centering
	\includegraphics[width = \columnwidth]{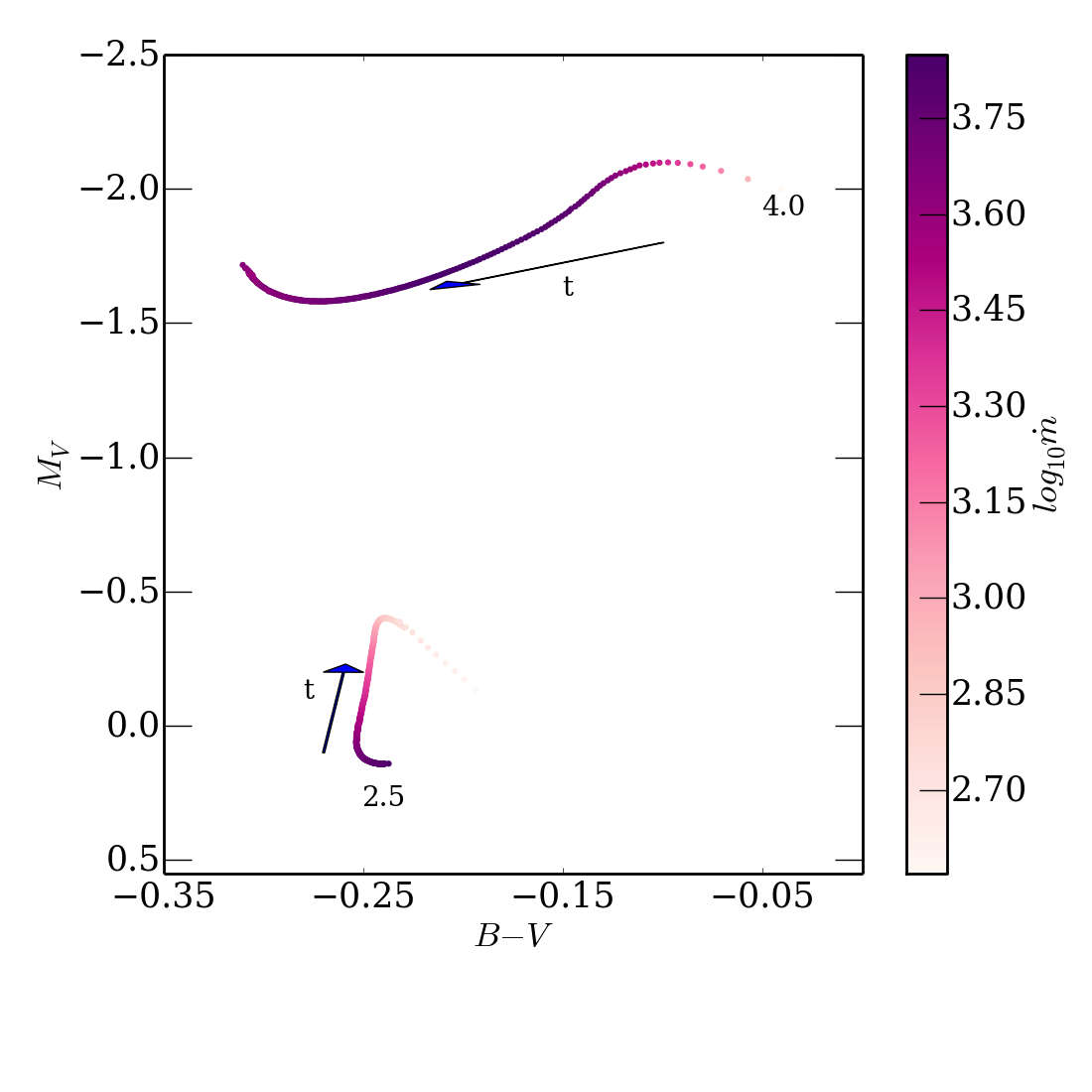}\caption{Colour-magnitude diagram in the $B-V$ colour and $M_{V}$ magnitude for systems with donors of 2.5 (lower line) and 4 (upper line) \msol accreting onto a NS with initial mass of 1.4 \msol . The colormap represents the mass transfer rate normalized to the Eddington limit. The arrows indicate the direction of the evolution.}\label{fig:cmd_ns}
\end{figure}
The evolutionary tracks on the CMD of the two binary pulsar ULXs described above are shown in Fig.~\ref{fig:cmd_ns}, where the colormap represents the magnitude of the mass transfer rate normalized to the Eddington limit of the evolving Neutron Star. \\
Both tracks have blue colors but the V band magnitude is very faint and considerably weaker than that of the evolutionary tracks of BH ULXs. The optical counterparts of ULX pulsars are essentially undetectable with HST if the masses of the donor stars are in the range 2.5-4 \msol. This result is consistent with the non-detection of the counterpart of NGC 5907 ULX-1 \citep{2013MNRAS.434.1702S,Heida_2019_donorsULP}. However, systems like NGC 7793 P13 that have a massive companion of 18-21 \msol in an eccentric orbit \citep{2017MNRAS.466L..48I,2017ApJ...835L...9H} can not clearly be reproduced  by a transient accretion phase of the type considered here. 


%

%
%
%
%
%
%
\section{Summary and Conclusions}
We calculated an extended grid of evolutionary tracks of ULX binary systems with the MESA code. We considered a wide range of BH masses (10-100 \msol) and donor masses (5-30 \msol) and properly incorporated the orbital angular momentum loss in case of super-Eddington accretion with an outflow. For the calculation of the tracks we adopted the model described in AZ.
We found that they occupy two regions on the CMD, depending on the evolutionary stage of the donor.\\
When the donor is on the MS and accretion is sub-critical or marginally super-critical, the tracks are blue and their $M_{V}$ magnitude is limited to $\sim -6$. They occupy the bluer and comparatively fainter corner of the CMD.
We noticed that they are similar in shape for different BH masses, being the main difference the maximum value of $M_{V}$: more massive BHs generate more luminous tracks.\\
When the donor is ascending towards the Giant Branch, the properties of the evolutionary tracks are governed by the mass transfer rate, which is now super critical.
Moreover, the binary separation increases rapidly producing a more extended accretion disc, which emits a huge amount of flux in the optical band, that can be enhanced by self-irradiation.
This evolution drives the tracks towards higher V band luminosities. The colors are initially blue because the mass transfer rate at the beginning of the Giant phase is very high and the disc is hot. As the super-Eddington mass transfer continues, the system widens and the accretion disc becomes more extended, driving the tracks towards redder colors. In addition, self-irradiation is essentially suppressed. However, after the peak, the mass transfer rate starts to decrease, irradiation starts to contribute again and for systems with more massive donors the evolution towards the red slows down. \\
In order to model the multiwavelength properties of ULXs we also included the contribution of saturated Comptonization from an inner spherical corona.\\
The data/model comparison proceeds through three sequential steps: comparison of the photometry with the evolutionary tracks on the CMD diagram, age selection, search for the best fit of the multiwavelength SED.\\
We applied our model to a number of ULXs with well studied stellar optical counterparts: NGC 4559 X$-$7, NGC 5204 X$-$1, HolmbergII X$-$1 and NGC 5907 ULX$-$2. 
\begin{itemize}

\item[-] \textit{NGC 4559 X-7: } The optical photometry of this source, together with the X-ray data, point to super-Eddington accretion onto massive BHs for this system. We found that the observed properties are reproduced by a system accreting above Eddington ($\dot{m}: \sim 900-1000$) onto a BH of $\sim$ 55 \msol from a donor of $\sim$ 5.5 \msol. The system consists of a standard disc emitting most of the optical luminosity,  an optically thick and extended outflow which covers part of the standard disc and almost the entire inner advection dominated disc, and an optically thick cool corona, with electron temperature of $\sim 1.2 $ keV, which dominates the hard X-ray emission. The orbital period is fairly large, $\sim 18$ d. In fact, when accretion occurs at super-critical rates, the orbital separation increases rapidly.  The system produces very high bolometric luminosity ( $(3-4)\times10^{40}$ erg s$^{-1}$ ), which is in agreement with the observed data. The results obtained with the MESA tracks are in agreement with those obtained with the Eggleton evolutionary tracks (see AZ). This source was studied previously by other authors, and our results are in agreement with their findings. PZ found that the optical counterpart of X-7 is reproduced by a massive BH which accretes from a donor of 30-50 \msol during the H-shell burning phase. \cite{2006MNRAS.368..397S} analyzed some \textit{XMM-Newton} observations of this source (which they refer to as NGC4559 X-1) finding that its spectral shape is consistent with super-Eddington accretion onto BHs with masses up to 80 \msol.

However, the estimated value of the BH mass inferred from our model is strongly dependent on the assumed distance of the source, which affects the optical luminosity of the system and hence the match with the evolutionary tracks on the CMD. The distance of NGC 4559 is quite uncertain and estimates vary in the range $\sim 6 - 15$ Mpc. For a distance lower than the one assumed here (10 Mpc), like the redshift-independent distance adopted by \cite{2018A&A...609A..37C} (d = 7.14 Mpc), lower mass BHs, down to 20 \msol, would be inferred. In a recent work, \cite{2021MNRAS.tmp..933P}  suggest the possibility that the compact object in NGC 4559 X-7 is even lighter, with a mass in the range 1.4 - 4 \msol (a light BH or a NS). A dedicated investigation of this scenario is outside the goals of this paper and we plan to do it in the future.

\item[-] \textit{NGC 5204 X-1: } The result obtained from the $\chi ^{2}$ fit of the SED at the time of intersection on the CMD suggests accretion at marginally super-critical rates from a donor with mass of $\sim$ 14 \msol onto a BH with mass of $\sim 40$ \msol . The system is $6.8$ Myrs old and accretion is taking place while the donor, whose initial mass is $\sim$ 30 \msol, is reaching the TAMS. However, the fit is not statistically acceptable and the bolometric luminosity is not well reproduced. The reason for which the fit is not satisfactory is related to the fact that the hard X-ray spectrum is not well represented. The source is harder than what predicted by our model and seems to require an additional high-energy tail to fit the X-ray data or a different geometry for comptonizing material. 
The existence of such a tail was studied by other authors. \cite{2015ApJ...808...64M} modelled it  with an additional Comptonizing component on the basis of the spectral analysis of coordinated \textit{XMM} and \textit{NuSTAR} observations of NGC 5204 X-1. A high energy excess with respect to a model with two thermal components has been found also by \cite{2018MNRAS.473.4360W}, and was modelled with a cut-off power law component, produced by a possible accretion column, should the compact object being a NS.

\item[-] \textit{HolmbergII X-1: } The modelling of this source shows that this system is accreting at marginally super-Eddington rates. The BH may have a mass of about 55 	\msol while the donor star has $\sim 11$ \msol . The best fit of the SED returns a corona temperature $ \sim 1 $ keV.  \cite{2012MNRAS.422..990K} studied the evolution of the spectral curvature of HolmbII finding that the accretion disk with advection (\textit{diskpbb}\footnote{In the X-ray spectral fitting package XSPEC the \textit{diskpbb} model is a multi-temperature blackbody disc model, where the local disc temperature T(r) is proportional to $ r^{-p}$, where p is a free parameter (see \citealt{1994ApJ...426..308M}).} with $p \approx 0.5$ in XSPEC) fits the data better than the standard accretion disc (\textit{diskbb}\footnote{In the X-ray spectral fitting package XSPEC the \textit{diskbb} model refers to a multi-temperature blackbody disc model, where the local disc temperature T(r) is proportional to $r^{-3/4}$ (see \citealt{1984PASJ...36..741M}).}).
Moreover, adding a Comptonizing component reproduces the spectrum well if a low electron temperature is considered ($kT_{e} \approx 1.2 $ keV). Their findings are in agreement with the results obtained with our model.

\item[-] \textit{NGC 5907 ULX-2:} The observed multiwavelength SED is better reproduced with a system composed of a donor star with mass of about 4.5 \msol which is accreting at super-Eddington rates onto a BH of about 36.5 \msol according to the value suggested by \cite{2018MNRAS.477L..90P} (where they adopted the model in AZ to infer a BH mass of 30 \msol). The system initially consisted of a 15 \msol donor and a 30 \msol BH. The accretion phase takes place while the donor is ascending along the Giant Branch. The binary has an orbital period of $\sim 24$ d, therefore the disc is very extended.
As reported by \cite{2018MNRAS.477L..90P} the color and luminosity of the optical counterpart are consistent with those of an O-B type star. From our modelling, it is clear that the contribution of the accretion disc to the optical emission is dominant.
Therefore the overall scenario can be misinterpreted if optical emission is ascribed only to the donor star. The outer standard disc contributes substantially to the optical flux mimicking the emission of a massive donor.
\end{itemize} 
In summary, we found that two out of four ULXs are well represented by systems with BH mass in the range between 35-40 \msol , while for the other two the best-fit solution points to more massive BHs, $\approx 55$\msol. 
Therefore, although the present sample of ULXs is very small and not representative of the whole population, the observational emission properties, together with binary evolution and age estimates, seem to suggest the existence of massive BHs in some sources.\\
We finally explored the possibility to extend our model to the challenging case of PULXs. We could steadily evolve only systems with a maximum donor mass of 5 \msol , whose optical emission results to be too faint to be detected with HST at the observed distances of ULXs. This result is consistent with the non-detection of the counterpart of the Pulsar ULX NGC 5907 ULX-1 \citep{2013MNRAS.434.1702S}. \\
We note that two PULXs have a detected optical counterpart: NGC 7793 P13 and NGC 1313 X-2 \citep{2008A&A...486..151G,2014Natur.514..198M,2019MNRAS.488L..35S}. The first one is accreting from a massive donor (18-23 \msol) \citep{2017MNRAS.466L..48I} and has been proved to be an eccentric binary system \citep{2017MNRAS.466L..48I,2017ApJ...835L...9H}. The optical counterpart of NGC 1313 X-2 is very luminous ($M_{V;B} < -4.5$) and cannot be reproduced with our evolutionary tracks for PULXs. Moreover, population studies reveal that its host environment is a young O-B association, aged $\approx 20 \, \text{Myr}$, pointing to a young massive star for the donor. If compared with our model of accreting BHs, the optical counterpart of X-2 would be represented by a 25 \msol BH which accretes at marginally super-Eddington rates from a 12 \msol donor . The apparent degeneracy intrinsic to this result is broken by the value of the orbital period (see also AZ), which would be too long to be consistent with the range of periods proposed by \cite{2019MNRAS.488L..35S}.\\ 
Related to this, we emphasize the importance of a multi-wavelength and multi-disciplinary approach to unveil the nature of the binary systems that power ULXs. This would help to discern the nature of the accretor whereas the possibility to detect coherent pulsation is halted by the poor quality of the data and/or the geometry of the system with respect to the observer or by the fact that pulsed emission is in fact not present. 
Despite the difficulties encountered by us and several authors in modelling accreting binaries with massive donors and a pulsars, these two sources show that accretion from a massive donor onto a neutron star is feasible.
Possible scenarios could be different from that proposed in this work. \cite{2019A&A...628A..19Q} shows that nuclear timescale mass transfer is feasible for neutron star accretors with a different prescription for hydrogen and helium content of the donor star. \cite{2019A&A...622L...3E} invokes accretion driven by Wind Roche lobe overflow. Another possibility is accretion under the hypothesis of eccentric orbits.  All these scenarios need to be investigated in detail in the future in a framework similar to the one proposed here to asses the formation routes of NS-massive donors systems.
%
%
\section*{Acknowledgements}
The Authors thank the anonymous referee for Their useful comments.
EA acknowledges funding from the Italian Space Agency, contract ASI/INAF n. I/004/11/4. L.Z. also acknowledges financial support from the ASI/INAF grants 2017-14-H.0 and I/037/12/0, and from INAF grant "Sostegno alla ricerca scientifica main stream" (INAF Presidential Decree 43/2018). A.W. also aknowledges financial support  from the ASI- INAF grant n.2017-14-H.0. Simulations in this paper made use of the MESA code, the Python language and the packages Matplotlib and MESA-reader. We also aknowledge the use of public data from the HEASARC Archive and the Xspec spectral fitting package.

\section*{Data Availability }
All of the data underlying this article are already publicly available from ESA’s XMM-Newton Science Archive
(\url{https://www.cosmos.esa.int/web/xmm-newton/xsa}) and  NASA’s
HEASARC archive (\url{https://heasarc.gsfc.nasa.gov/}).

%
%
%



\bibliographystyle{mnras}
\bibliography{biblio_multiw} 



\appendix
\section{Z = 0.2 $Z_{\odot}$, BH = 20 \msol}\label{app:metallicity}
Effects of metallicity on the binary evolution of systems that undergo case A mass transfer. We show the representative case of accretion onto a BH of 20 \msol for donors with solar and sub-solar (Z = 0.2 $Z_{\odot}$) metallicity. Metallicity does not induce important changes on the evolution of donors for case A mass transfer. 
\begin{figure*}\label{fig:diffz}
\includegraphics[width=\textwidth]{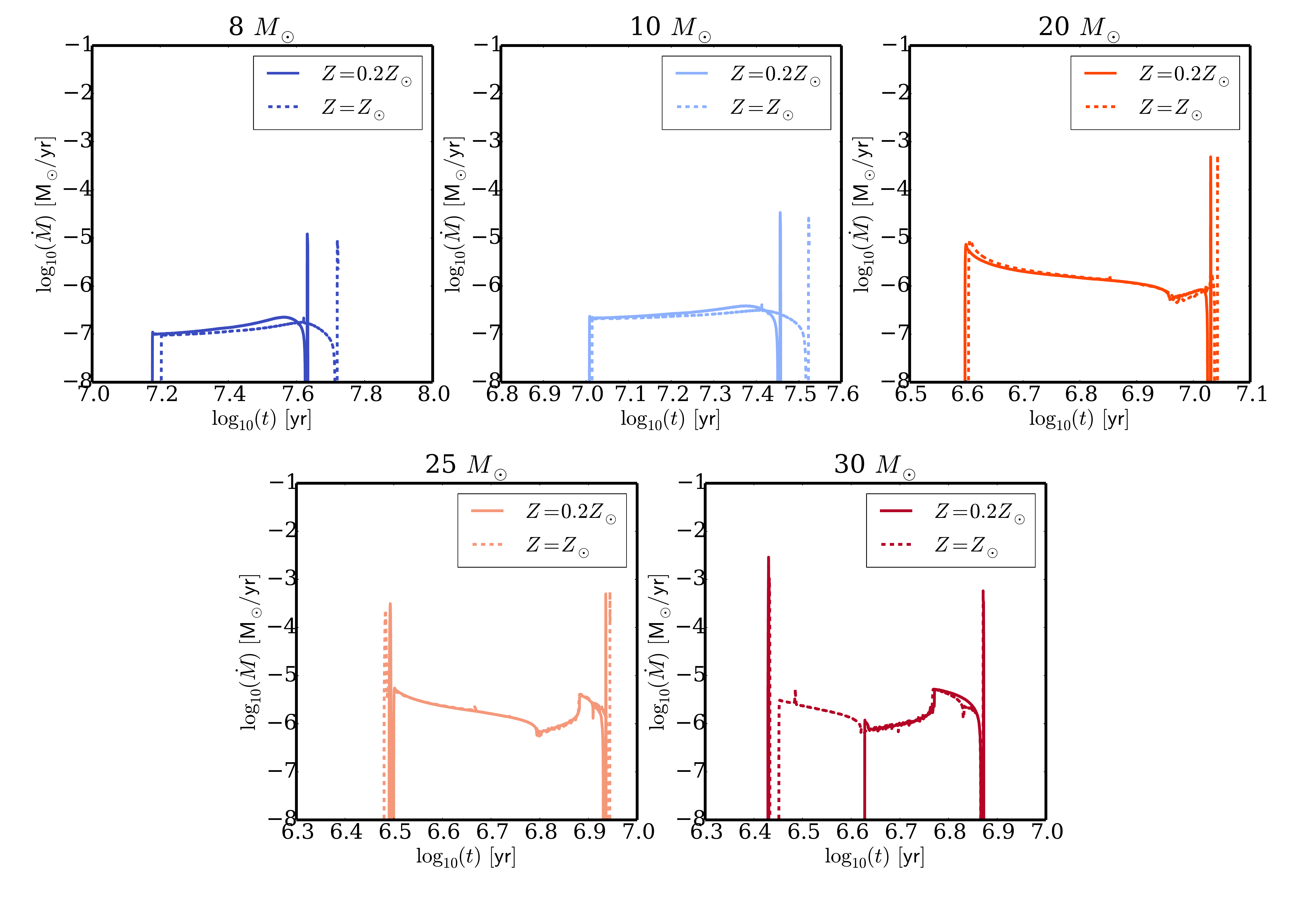}\caption{Evolutionary tracks for systems with same donor mass and BH mass, but different metallicities.}
\end{figure*}
Fig. \ref{fig:diffz} shows the evolution of the mass transfer rate for systems with donor with 8, 10, 20, 25 and 30 \msol . The mass transfer rate evolves similarly in the two cases, with a slight delay in time and shortening in the duration of the main accretion episode.
\section{Evolutionary tracks for a given BH mass}
\label{app:systems}
\subsubsection*{Systems accreting onto a 10 \msol BH}
The evolutionary tracks on the CMD for systems accreting onto a 10 \msol BH are shown in the upper left panel of Figs.~\ref{fig_CMD_MESA_bhfixed} and ~\ref{fig:CMD_MESA_bhfixed_vmeni}.
We notice first that the track of the 20 \msol donor does not appear in the CMD, while that of a 15 \msol donor is only marginally represented. During the first contact episode during MS, these systems accrete at very high super-Eddington rates (see the upper left panel of Fig.~\ref{fig:mbhconst_mdot}) so that the outflow engulfs the binary and we cannot follow their evolution with our model.  We are left with the systems accreting from less massive donors: from 5 to 12 \msol .\\ When the donors are on the MS, the tracks are blue ($B-V \simeq -0.3 : 0.0 $) and not very luminous ($M_{V} \simeq -2 : -3.5 $). The 5 \msol donor represents an exception (see below): during Main Sequence it becomes redder ($B-V$ increases up to 0.5) and its luminosity progressively decreases ($M_{V} \simeq -1 : -2.3 $).\\
When the donors evolve off the MS and the mass transfer becomes super-Eddington, colors are initially blue and then they become redder. The magnitudes are brighter than in MS, in the range $M_{V} = -3.8 : -5.2 $.\\
As mentioned above, the 5 \msol donor behaves in a different way. 
First, the radius of the disc at the TAMS is larger than that of the other systems. Moreover, the mass transfer rate is considerably sub-Eddington. 
These two facts imply that the accretion disc is relatively cold. Also the donor star, which at TAMS is a low mass star (see Fig.~\ref{fig:parameters}), is relatively weak and cold. 
Therefore, the V magnitude of the system is weak and colors are red ($B-V \sim 0.5$). Given the low star and disc temperatures, the system with a 5 \msol donor is more luminous in the I band than in the V band, as it can be seen comparing Fig. \ref{fig_CMD_MESA_bhfixed} with Fig.~\ref{fig:CMD_MESA_bhfixed_vmeni}.
\subsubsection*{Systems accreting onto a 20 \msol BH}
Let now consider the CMDs for systems accreting onto a 20 \msol BH (upper right panel of Figs.~\ref{fig_CMD_MESA_bhfixed} and \ref{fig:CMD_MESA_bhfixed_vmeni}). At variance with the previous case, the tracks belonging to all the donor stars appear in the diagram. In fact, the lower value of the mass ratios (Fig.~\ref{fig:parameters}) allows for a less eruptive initial mass transfer phase.
We could evolve steadily the 15 \msol donor.  The track of a 20 \msol donor is stopped only at the final stages of the giant branch. More massive donors (25 and 30 \msol) are evolved for a short time and then their evolution is stopped because the outflow engulfs the binary.\\
The evolution of the optical properties of these systems are similar to those described above for a 10 \msol BH.
%
The only difference is that V-I colors are redder than the those of systems with a 10 \msol BH, because the discs are more extended 
The evolution of the 5 \msol  donor during MS is different, because the system is accreting sub-Eddington (as for the 10 \msol BH).\\
For the more massive donors ($M>10 M_{\odot}$), the accretion in the post MS phases is characterized by blue colors and bright $M_{V}$ magnitude ($B-V \sim -0.23 : 0.0 $, $M_{V} \sim -5.0 : -6.3$). For lower mass donors ($M \leq 10 M_{\odot}$), systems are comparatively redder and fainter ($B-V \simeq 0 : 0.6 $, $V-I \sim -0.2 : 0.8$, $M_{V} \sim -4.0 : -5.3$ , $M_{I} \sim -4.5 : -6.1$). 
\subsubsection*{Systems accreting onto a 30 \msol BH}
For a 30 \msol BH accretion is more stable from the beginning of the RLOF for donors up to 30 \msol (middle left panel of Fig.~\ref{fig:mbhconst_mdot}). For a given mass the binary separation is larger. This prevents the outflow from engulfing the binary during MS. All the systems with donor mass $M_{d} \leq 25$ \msol can be evolved with our model, while the system with donor mass $M_{d} = 30$ \msol is stopped at the middle of the giant phase. 
For donors less massive than 20 \msol during MS, the flux emitted by the disc overcomes that emitted by the star  (see Appendix ~\ref{app:fluxratios}). The evolution of the optical emission is essentially similar to that of the 10 and 20 \msol BH case.\\
%
%
When systems exit the MS the flux of the disc is dominant and the emission is super-Eddington. The tracks of more massive donors remain blue, and luminous ($B-V \sim -0.0 : -0.05 $, $V-I \sim -0.4 : 0.1$, $M_{V} = -5.1 : -6.9 $, $ M_{I} \sim  -5.0 : -6.7$). Systems with less massive donors turn into redder colors and are slightly less luminous ($B-V \sim -0.1 : 0.4 $, $V-I \sim -0.3 : 0.9 $, $M_{V} = -1.5 : -5.5 $, $ M_{I} \sim  -4.5 : -6.5$).
\subsubsection*{Systems accreting onto a 50 \msol BH} 
For a 50 \msol BH (middle-right panel of Figs.~\ref{fig_CMD_MESA_bhfixed} and \ref{fig:CMD_MESA_bhfixed_vmeni}) only the track with the donor of 30 \msol is truncated when the outflow engulfs the binary.
%
%
The post-MS phases, characterized also in this case by super-Eddington accretion, follow the trend described above.
The range of colors and magnitudes of these systems during the super-Eddington accretion phases are:  $B-V \sim  -0.1: 0.4 $, $V-I \sim -0.4  : 1.0 $, $M_{V} =-5.2 : -7.5 $, $ M_{I} \sim -5.4 : -7.2$
\subsubsection*{Systems accreting onto a 70 \msol BH}
For a 70 \msol BH (lower left panel of Figs.~\ref{fig_CMD_MESA_bhfixed} and ~\ref{fig:CMD_MESA_bhfixed_vmeni}), the evolution of the systems is similar to that of the 50 \msol BH. Also in this case, the track of the 30 \msol donor is interrupted because the outflow becomes too extended.  
%
The tracks are similar to the 50 \msol case but slightly redder, being the accretion disc larger. Magnitude and colors during MS are in the range: $B-V \sim -0.3 : 0.6  $, $V-I \sim -0.4 : 0.5 $,  $M_{V} = -2.5 : -6.0 $, $ M_{I} \sim -2.8 : -5.9$, while during the post-MS phases are in the range: $B-V \sim -0.1 : 0.4  $, $V-I \sim -0.4 : 1.0 $ , $M_{V} = -5.5 : -7.8 $, $ M_{I} \sim -5.8 : -7.7$.
\subsubsection*{Systems accreting onto a 100 \msol BH}
Finally, we describe the evolution of systems accreting onto a 100 \msol BH, which is the maximum BH mass considered in this work (lower right panels in Figs.~\ref{fig_CMD_MESA_bhfixed} and ~\ref{fig:CMD_MESA_bhfixed_vmeni}).
Even for the high mass transfer rate which characterizes the initial evolution of a 30 \msol donor, see Fig.~\ref{fig:mbhconst_mdot}, the outflow does not engulf the system. We can evolve it up to the moment in which the donor starts He-burning in the nucleus, which is the endpoint of the evolutionary tracks we evolved with MESA.\\
%
%
The behaviour is similar to that of the systems with a 50 or 70 \msol BH.
The systems become very red and the luminosity in the V band increases significantly.
Magnitude and colors, during MS are in the range: $B-V \sim -0.3 : - 0.1$, $V-I \sim -0.45 : 0.5 $,  $M_{V} = -2.5 :-6.3 $, $ M_{I} \sim -2.3 : -6.1$, while during the post-MS phases are in the range: $B-V \sim -0.25 : -0.1   $, $V-I \sim -0.4 : 0.1 $ , $M_{V} =  -6.0 : -8.2$, $ M_{I} \sim -6.1 : -7.9$.

\section{flux ratios}\label{app:fluxratios}
In this Appendix we show the ratios between the flux emitted by the disc and that emitted by the star in the V band These figures help to understand the evolutionary tracks on the CMD described in Sec.~\ref{sec:mesa_tracks_cmd}.
\begin{figure*}
	\includegraphics[width=\columnwidth]{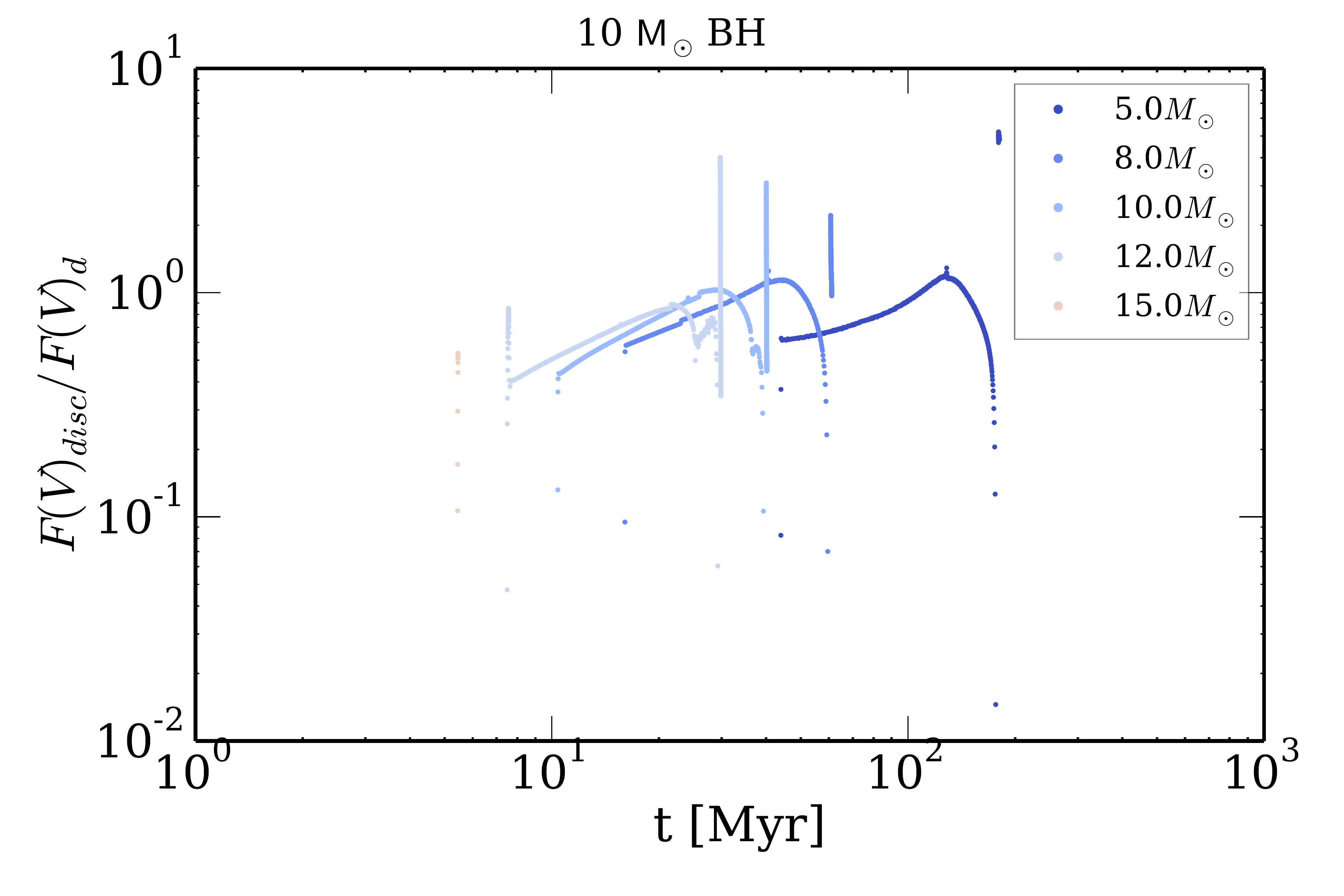}
	\includegraphics[width=\columnwidth]{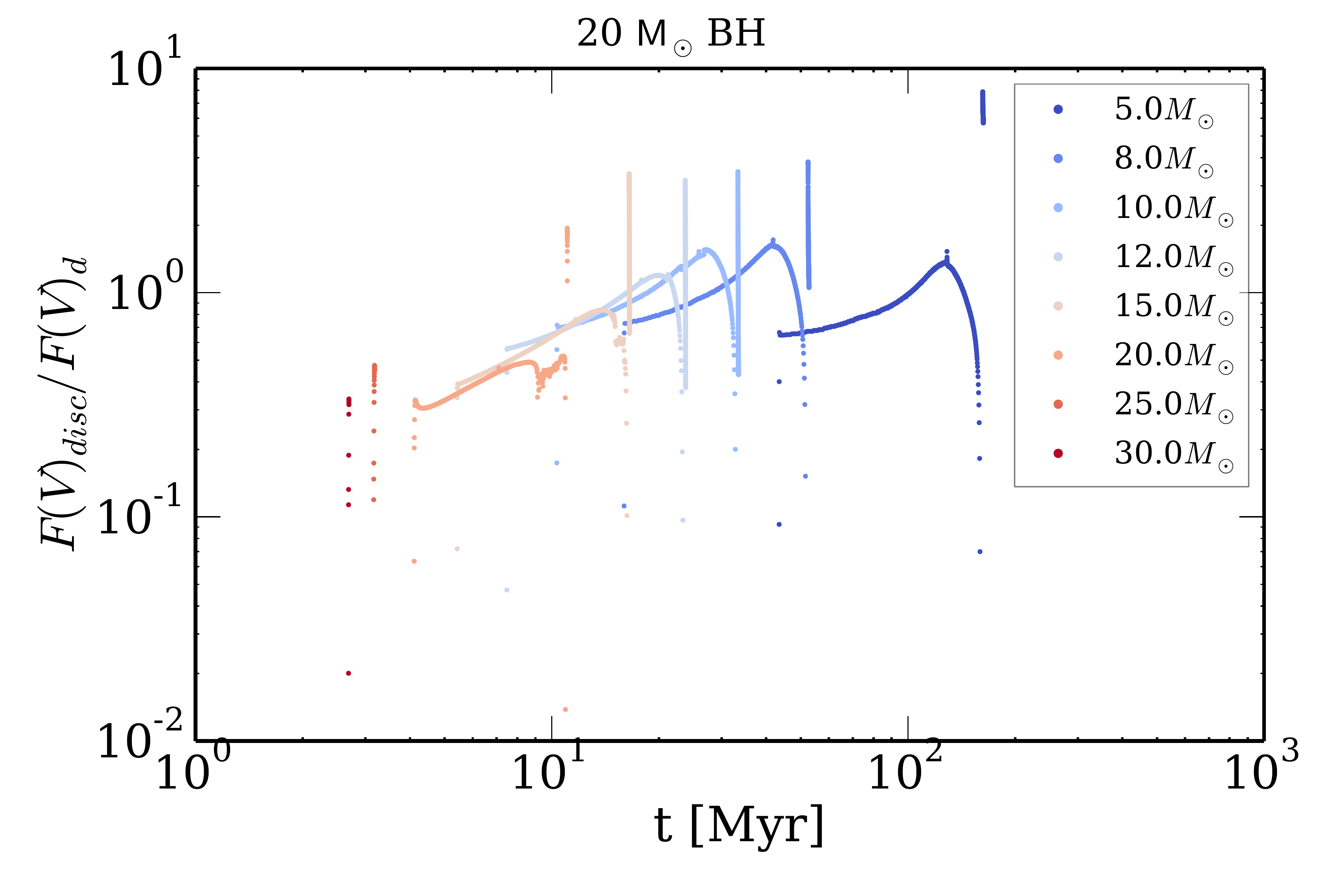}
	\includegraphics[width=\columnwidth]{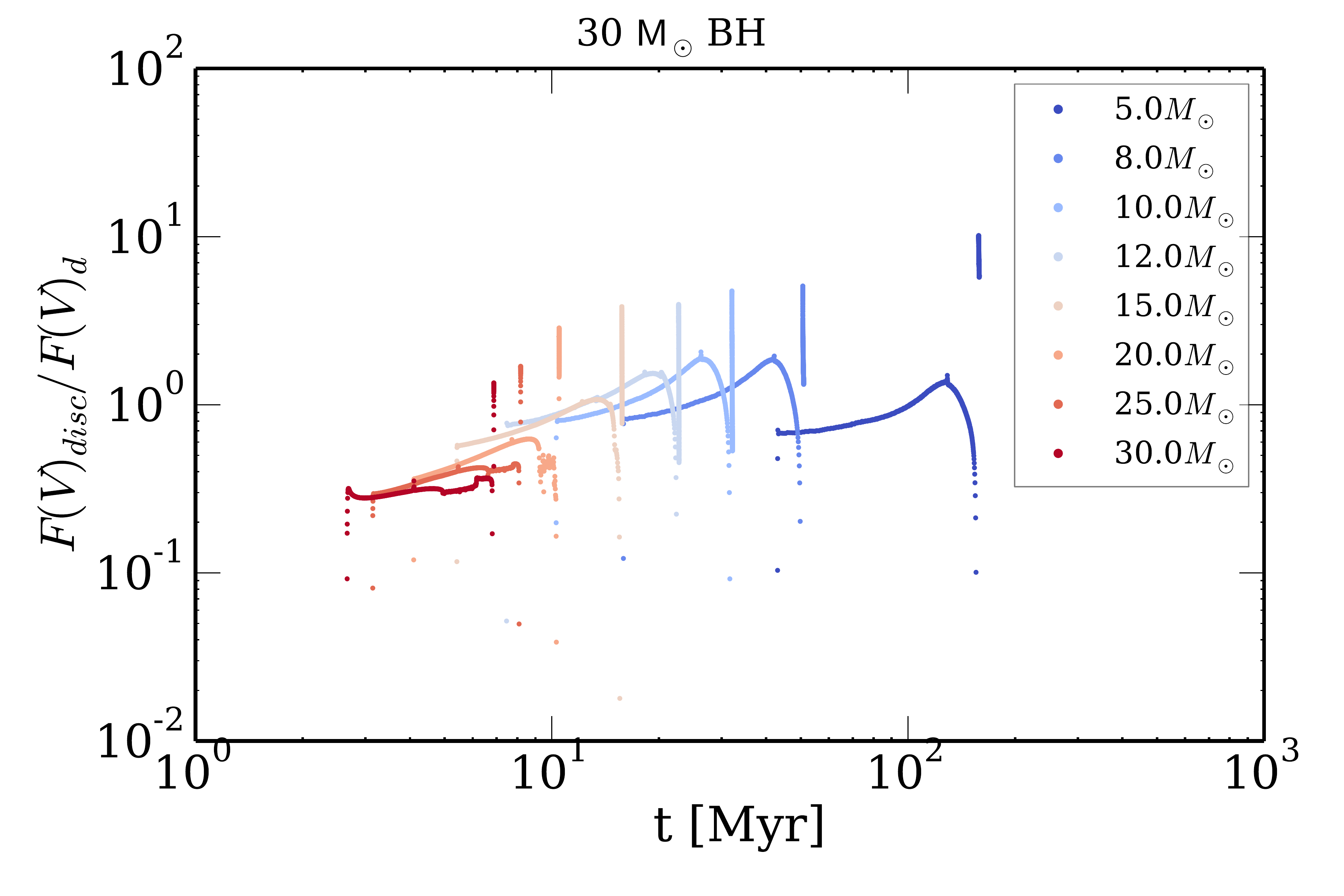}
	\includegraphics[width=\columnwidth]{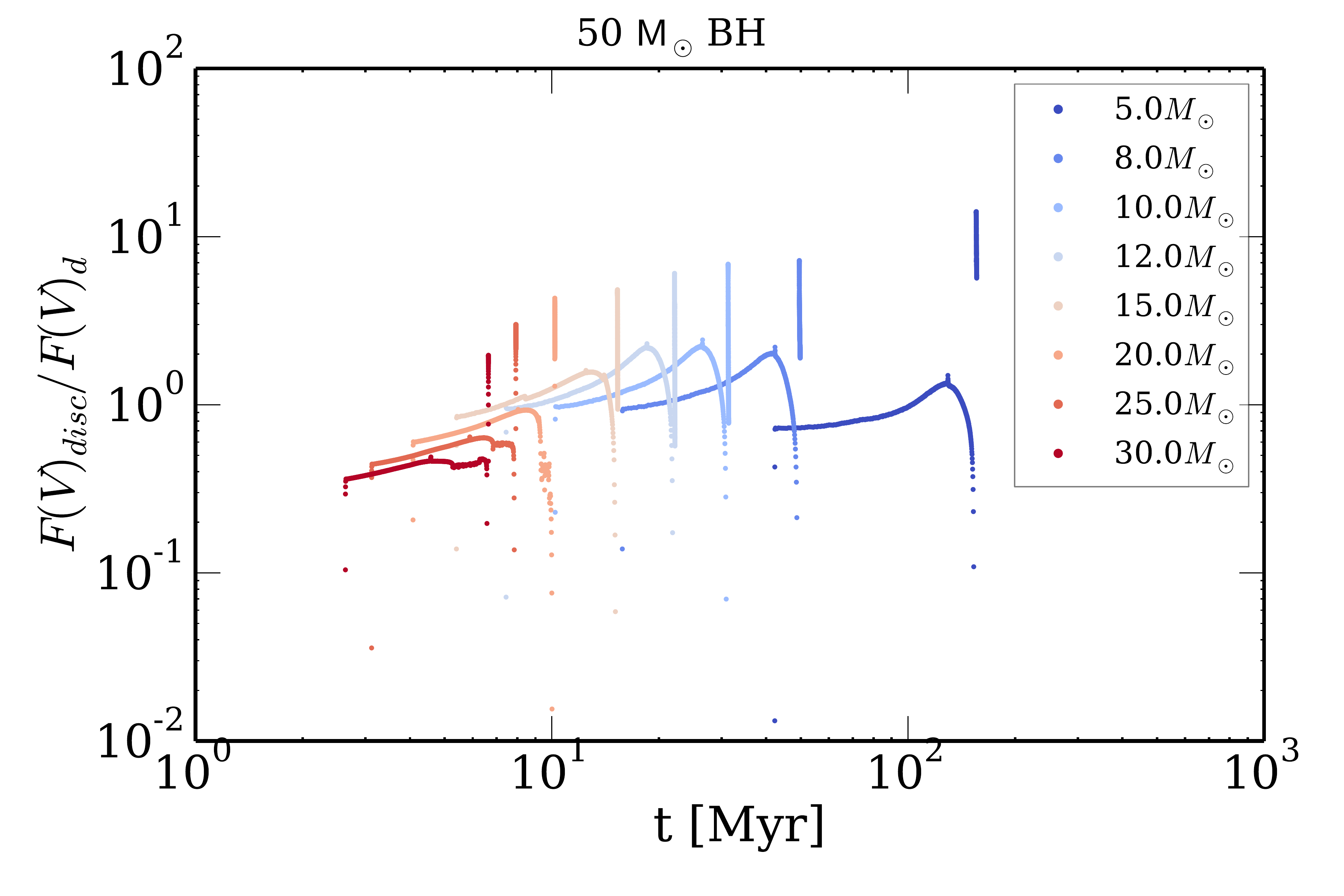}
	\includegraphics[width=\columnwidth]{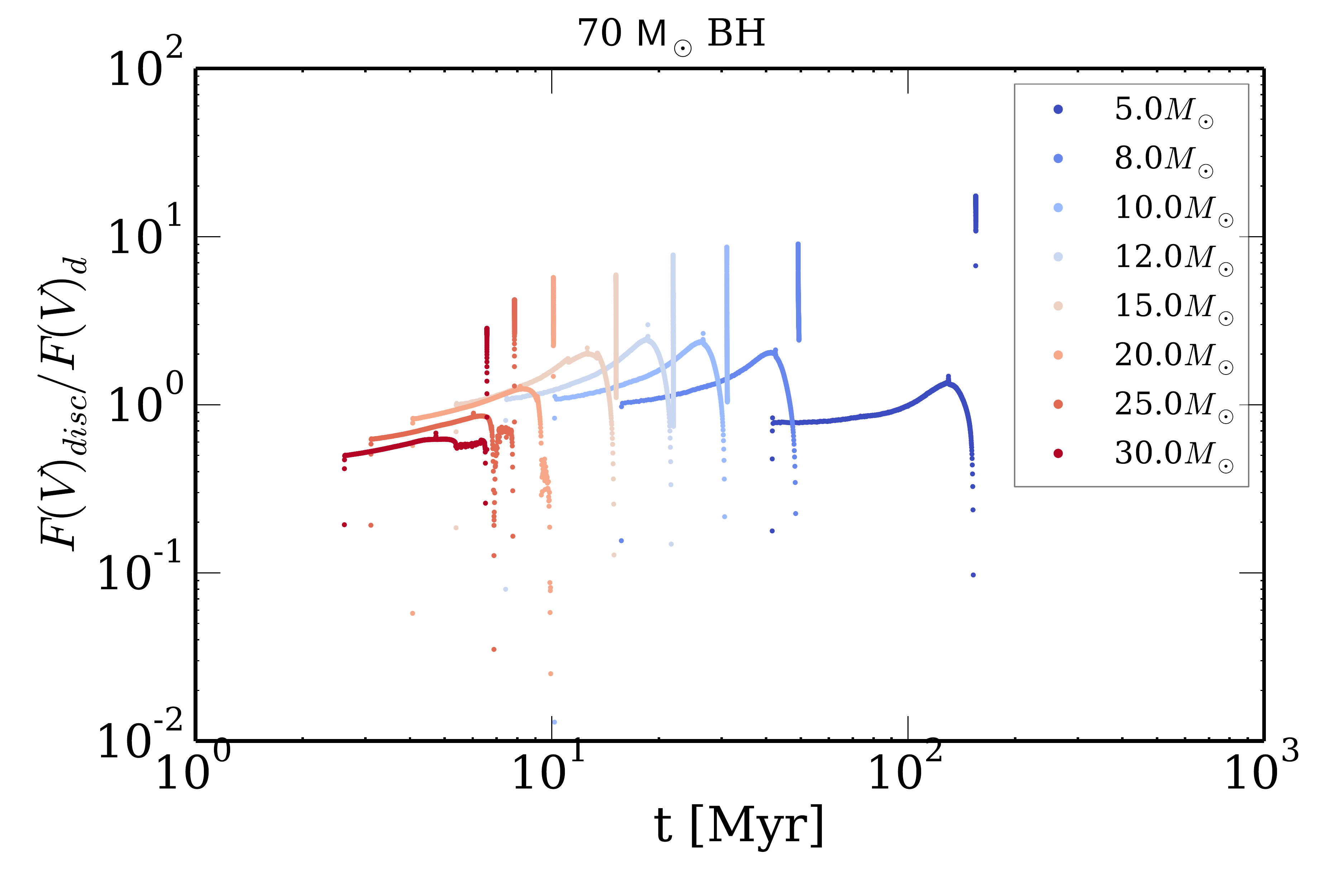}
	\includegraphics[width=\columnwidth]{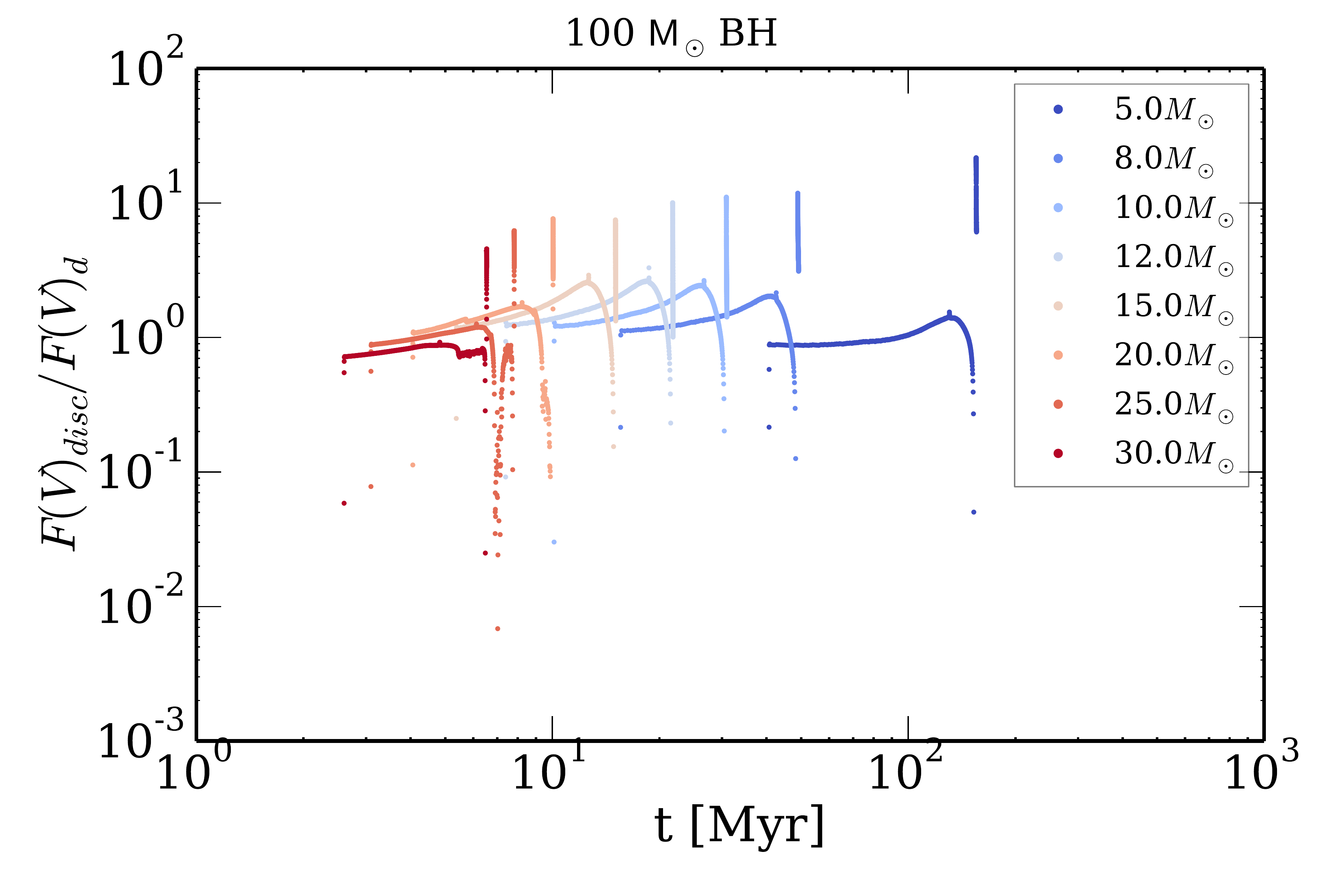}\caption{Evolution of the ratio between the flux emitted by the disc and that emitted by the donor star in the Johnson V filter. The trend of the tracks while the donor is on the Main Sequence is considerably influenced by the flux ratio. }
\end{figure*}
	


\bsp	
\label{lastpage}
\end{document}